\documentclass[12pt]{article}

\usepackage{graphicx,epstopdf,amsmath}

\def\be{\begin{equation}}
\def\ee{\end{equation}}
\def\ba{\begin{eqnarray}}
\def\ea{\end{eqnarray}}
\def\ge{\mathrel{\raise.3ex\hbox{$>$\kern-.75em\lower1ex\hbox{$\sim$}}}}
\def\la{\mathrel{\raise.3ex\hbox{$<$\kern-.75em\lower1ex\hbox{$\sim$}}}}

\def\simgt{\mathrel{\raise.3ex\hbox{$>$\kern-.75em\lower1ex\hbox{$\sim$}}}}
\def\simlt{\mathrel{\raise.3ex\hbox{$<$\kern-.75em\lower1ex\hbox{$\sim$}}}}

\newcommand{\bi}[1]{\bibitem{#1}}
\newcommand{\fr}[2]{\frac{#1}{#2}}

\newcommand{\Tr}{\mbox{Tr}}

\newcommand{\nc}{\newcommand}

\nc{\gone}{\bar g_{\pi NN}^{(1)}}
\nc{\gzero}{\bar g_{\pi NN}^{(0)}}
\nc{\al}{\alpha}
\nc{\ga}{\gamma}
\nc{\de}{\delta}
\nc{\ep}{\epsilon}
\nc{\ze}{\zeta}
\nc{\et}{\eta}
\nc{\ka}{\kappa}
\nc{\rh}{\rho}
\nc{\si}{\sigma}
\nc{\ta}{\tau}
\nc{\up}{\upsilon}
\nc{\ph}{\phi}
\nc{\ch}{\chi}
\nc{\ps}{\psi}
\nc{\om}{\omega}
\nc{\Ga}{\Gamma}
\nc{\De}{\Delta}
\nc{\La}{\Lambda}
\nc{\Si}{\Sigma}
\nc{\Up}{\Upsilon}
\nc{\Ph}{\Phi}
\nc{\Ps}{\Psi}
\nc{\Om}{\Omega}
\nc{\ptl}{\partial}
\nc{\del}{\nabla}
\nc{\ov}{\overline}
\nc{\newcaption}[1]{\centerline{\parbox{15cm}{\caption{#1}}}}

\def\beq{\begin{equation}}
\def\eeq{\end{equation}}
\def\bmat{\begin{displaymath}}
\def\emat{\end{displaymath}}
\def\bear{\begin{eqnarray}}
\def\eear{\end{eqnarray}}
\def\ba{\begin{eqnarray}}
\def\ea{\end{eqnarray}}
\def\bery{\begin{array}}
\def\ery{\end{array}}
\def\bit{\begin{itemize}}
\def\eit{\end{itemize}}
\def\ben{\begin{enumerate}}
\def\een{\end{enumerate}}
\def\btab{\begin{tabular}}
\def\etab{\end{tabular}}
\def\btbl{\begin{table}}
\def\etbl{\end{table}}
\def\bfig{\begin{figure}[htb]}
\def\efig{\end{figure}}
\def\bpic{\begin{picture}}
\def\epic{\end{picture}}

\def\ga{\mathrel{\raise.3ex\hbox{$>$\kern-.75em\lower1ex\hbox{$\sim$}}}}
\def\la{\mathrel{\raise.3ex\hbox{$<$\kern-.75em\lower1ex\hbox{$\sim$}}}}
\def\gappeq{\mathrel{\rlap {\raise.5ex\hbox{$>$}}
{\lower.5ex\hbox{$\sim$}}}}
\def\lappeq{\mathrel{\rlap{\raise.5ex\hbox{$<$}}
{\lower.5ex\hbox{$\sim$}}}}

\def\gyr{{\rm \, G\kern-0.125em yr}}
\def\mev{{\rm \, Me\kern-0.125em V}}
\def\gev{{\rm \, Ge\kern-0.125em V}}
\def\tev{{\rm \, Te\kern-0.125em V}}

\def\Tr{\rm Tr}

\begin{document}

\begin{titlepage}

\setcounter{page}{1}

\vspace*{0.2in}

\begin{center}

\hspace*{-0.6cm}\parbox{17.5cm}{\Large \bf \begin{center}
Sensitivity to new supersymmetric thresholds through flavour and $CP$ violating physics\end{center}}

\vspace*{0.5cm}
\normalsize

{\bf  Maxim Pospelov$^{\,(a,b)}$, 
Adam Ritz$^{\,(a)}$ and Yudi Santoso$^{\,(a)}$}

\smallskip
\medskip

$^{\,(a)}${\it Department of Physics and Astronomy, University of Victoria, \\
     Victoria, BC, V8P 1A1 Canada}

$^{\,(b)}${\it Perimeter Institute for Theoretical Physics, Waterloo,
ON, N2J 2W9, Canada}

\smallskip
\end{center}
\vskip0.2in

\centerline{\large\bf Abstract}

Treating the MSSM as an effective theory below a threshold scale $\La$, we study the consequences
of having dimension-five operators in the superpotential for flavour and $CP$-violating processes.
Below the supersymmetric threshold such terms generate flavour changing and/or $CP$-odd 
effective operators of dimension six composed from the Standard Model fermions,
that have the interesting property of decoupling linearly with the threshold scale, i.e. as 
$1/(\Lambda m_{\rm soft})$, where $m_{\rm soft}$ is the scale of soft supersymmetry breaking.
The assumption of weak-scale supersymmetry, together with the stringent limits on electric dipole moments 
and lepton flavour-violating processes, then provides sensitivity to $\Lambda$ as high as 
$10^7-10^9$ GeV. We discuss the varying sensitivity to these scales within several MSSM benchmark scenarios
and also outline the classes of UV physics which could generate these operators.

\vfil
\leftline{August 2006}

\end{titlepage}

\section{Introduction}

Weak-scale supersymmetry (SUSY) is a theoretical framework 
that helps to soften the so-called gauge hierarchy problem 
by removing the power-like sensitivity of the dimensionful parameters in the 
Higgs potential to the square of the ultraviolet cutoff $\Lambda$. 
This feature, among others, has stimulated a large body of theoretical work on weak-scale supersymmetry,
supplemented by continuing experimental searches, which now spans almost 
three decades. Yet the supersymmetrized version of the Standard Model (SM), the minimal supersymmetric 
Standard Model (MSSM), suffers from well known problems such as the large array of 
allowed free parameters responsible for soft SUSY breaking, and the consequent
possibility of large flavour and $CP$ violating amplitudes. 
The absence of $CP$-violation at the ${\cal O}(1)$ level in the soft-breaking sector of 
the MSSM, as suggested by the null results of electric dipole moment (EDM) searches 
and the perfect accord of the observed $K$ and $B$ meson mixing and decay
with the predictions of the SM, implies that the 
soft-breaking sector of the MSSM somehow conserves $CP$ and does not 
source new flavour-changing processes. Whether or not such a pattern of soft-breaking masses 
is theoretically feasible is the subject of on-going studies addressing the mechanism of SUSY breaking 
and mediation (see, {\it e.g.}  \cite{mssm}). In this work, we will make the assumption that an (approximately) 
flavour-universal and $CP$-conserving soft-breaking sector is realized, and study the consequences of 
the presence of SUSY-{\em preserving} higher-dimensional operators on flavour and $CP$-violating 
observables. 

These operators may be thought to emerge from new physics at some 
high-energy scale $\Lambda$, which is larger than the electroweak scale. 
Even though the field content of the MSSM may be perfectly `complete' 
at the electroweak scale, it is clear that almost by construction the MSSM cannot be 
a fundamental theory because of the required high-energy 
physics responsible for SUSY breaking and mediation. In recent years there is also a more phenomenological 
motivation for a new threshold, namely the new physics responsible for neutrino masses (assuming they
are Majorana) and mixings.  Beyond these primary concerns, the possibility of new thresholds, intermediate between the 
weak and the GUT scales, is also suggested by the axion solution to the strong $CP$ problem, by the  SUSY
leptogenesis scenarios \cite{leptogen} and, more entertainingly, by the possibility of a lowered 
GUT/string scale arising from the large radius compactification of extra dimensions \cite{LED}.
In summary, given the assumed existence of weak-scale supersymmetry, there seems ample motivation to
expect additional new physics thresholds above the electroweak scale and possibly below the GUT scale. 
The presence of such thresholds will generically be manifest not just through corrections to relevant and marginal
operators, but also through the presence of higher-dimensional operators.

As is easy to see, both K\"ahler terms and the superpotential can receive additional 
non-renormalizable terms at the leading dimension five level \cite{Weinberg,SY}. Some of these operators are well-known
and were studied in connection with baryon-number violating processes and also the see-saw mechanism for neutrino masses.
However, to the best of our knowledge,  an analysis of the full set of dimension-five operators  with respect to flavour and $CP$-violating 
observables is still lacking. The purpose of this paper is thus to consider all possible dimension five extensions of the 
MSSM superpotential and K\"ahler terms, concentrating on those that conserve 
lepton and baryon number  and are $R$-parity symmetric.  We initiated such a study recently \cite{prs}, and will provide further details and 
extensions in the present work. As we shall see, such operators can induce 
large corrections to flavour-changing and/or $CP$-violating amplitudes and therefore 
can be efficiently probed with existing experiments and future searches. 

There is a clear parametric distinction between the effects induced by nonuniversal soft-breaking terms
and by the higher-dimensional extensions of the superpotential. Whereas the former typically 
scale as $m_{\rm soft}^{-2}$ times one or two powers of the flavour-mixing 
angle $\delta_{ij}$ in the squark(slepton) sector, the latter 
decouple as $(\Lambda m_{\rm soft})^{-1}\delta'_{ij}$, where  $\delta'_{ij}$ parametrizes flavour violation in 
the dimension five operators. When $\La$ is relatively large, and thus the threshold corrections to the soft-terms
are small, we may have scenarios where  
$\delta_{ij} \simeq 0$ while $\delta'_{ij}$ are significant, 
and the corrections to the superpotential can be the dominant mechanism for SUSY
flavour and $CP$ violation, providing considerable sensitivity to $\Lambda$.
At the same time, the additional $CP$ and flavour violation introduced in this way
 can be rendered harmless by simply increasing $\Lambda$. 

The layout of the paper is as follows. In the next section we list the possible 
operators in the MSSM superpotential and the K\"ahler terms 
at dimension five level, including for completeness those that violate $R$-parity. The relevant supersymmetric renormalization
group equations for the operators of interest are included in an Appendix. In section 3, we 
perform the required calculations at the SUSY threshold to connect this extension of the superpotential 
with the resulting Wilson coefficients in front of various effective SM operators of phenomenological interest. Section 4 
addresses the consequent predictions for the most sensitive $CP$-odd and 
flavour-violating amplitudes and infers the characteristic sensitivity to $\Lambda$ in each channel. In section 5, we perform this analysis
within four SPS benchmark scenarios~\cite{sps} (see also~\cite{bregop}) in
order to infer the dependence of this sensitivity on the features of the SUSY spectrum.
Section 6 contains a discussion and also a brief analysis of the general classes of new physics which could be responsible for
these operators, while our conclusions are summarized in section 7.

\section{Dimension-5 operators in the MSSM}

In this section, we will enumerate all the allowed structures in the 
superpotential and K\"ahler potential up to dimension 5 according to the standard symmetries 
of the MSSM (see, {\em e.g.} \cite{Weinberg,SY}).
We begin by recalling in Table~1 the  chiral superfields of the MSSM \cite{DG}
along with their gauge quantum numbers.
\begin{table}
\begin{center}
\btab{||c|c|c|c|c||}
\hline
Superfield & $SU(3)_C$ & $SU(2)_L$ & $U(1)_Y$ & $P_M$ \\
\hline
$Q$ & {\bf 3} & {\bf 2} & 1/6 & -1 \\
$U$ & $\overline{\bf 3}$ & {\bf 1} & -2/3 & -1 \\
$D$ & $\overline{\bf 3}$ & {\bf 1} & 1/3 & -1 \\
$L$ & {\bf 1} & {\bf 2} & -1/2 & -1 \\
$E$ & {\bf 1} & {\bf 1} & 1 & -1 \\
$H_u$ & {\bf 1} & {\bf 2} & 1/2 & +1 \\
$H_d$ & {\bf 1} & {\bf 2} & -1/2 & +1 \\ 
\hline
\etab
\caption{\footnotesize Representations and quantum numbers for chiral fields in the MSSM.}
\end{center}
\end{table}

The matter parity, $P_M$, is defined in the usual way,
\beq
P_M \equiv (-1)^{3(B-L)}
\eeq
where $B$ is the baryon number and $L$ the lepton number. 
This can be restated as $R$-parity, defined as
\beq
P_R = (-1)^{3(B-L) + 2s}
\eeq
where $s$ is the spin of the component field. All known Standard Model particles have
$P_R = +1$, while their superpartners have $P_R = -1$. However, when using the superfield
formalism it is often more convenient to use matter parity in which all fields
belonging to the same superfield have the same value of $P_M$. 

The superpotential of the MSSM contains a number of dimensionless parameters, and one 
dimensionful parameter $\mu$ or, in equivalent language, is composed from one 
dimension three and several dimension four operators:
\bear
{\cal W}^{(3)} &=& - \mu H_dH_u  
\\
{\cal W}^{(4)} &=& Y_u  U Q  H_u - Y_d D Q  H_d - Y_e E L H_d,
\eear
where gauge and generation indices are suppressed. 
All these terms conserve $R$-parity. In counting the dimensions, 
one should keep in mind that we are implicitly including dim$[d^2\theta] =1$.
At the renormalizable level, there are additional terms that
are forbidden by matter/$R$ parity but allowed by gauge invariance,
\bear
{\cal W}^{(3)}_{\not{R}} &=& -\mu' L H_u 
\\
{\cal W}^{(4)}_{\not{R}} &=& \lambda L L E +\lambda' LQ D  + \lambda^{''} UDD\, .
\eear

Going beyond the renormalizable level, at dimension-five there are a number of operators
allowed by symmetries. It is worth recalling that in the Standard Model, above the electroweak scale, there
is only a single class of dimension-five operators allowed by symmetries -- the seesaw operator -- which can naturally
provide a small Majorana mass for the active neutrinos. Within the MSSM, the list is only slightly longer. 
Suppressing a variety of  gauge and generational indices, the collection of 
dimension five operators can be presented in the following schematic form:
\bear
\label{dim5}
{\cal W}^{(5)} &=& c_{qq} Q U
 Q D + c_{qe} Q U L E + 
c_{h}H_u H_d H_u H_d\\ 
&& +c_{\nu}H_u L H_u L  + 
c_{p1}U U D E + c_{p2}QQQL\, . \nonumber 
\eear 
The final two terms in this list violate baryon and lepton number by 
one unit, and therefore induce proton decay. Detailed studies 
of these operators induced by triplet Higgs exchange have been 
conducted over the years in the context of SUSY GUT models \cite{Weinberg,SY} (for a recent 
assessment, see {\em e.g.} \cite{Murayama:2001ur}). 
The $H_u L H_u L$ operator is the superfield generalization of the SM see-saw operator and can be responsible 
for generating the neutrino masses and mixings. Assuming neutrinos are Majorana, the flavour structure
of $c_\nu$ is currently being determined in neutrino physics experiments (see {\em e.g.} \cite{neutrinos}). 

Going over to $R$-parity violating terms (see {\em e.g.} \cite{sakis}), 
one finds additional dimension five operators,
\beq
{\cal W}^{(5)}_{\not{R}} =  c^{\not{R}}_1Q U H_d E + c^{\not{R}}_2H_u H_d H_u L + c^{\not{R}}_3 QQQH_d,
\label{dim5rpv}
\eeq
that can be obtained from (\ref{dim5}) upon the simple substitution $L \leftrightarrow H_d$. 

If we now consider the K\"ahler potential, it is easy to see that at dimension four 
one has the standard $\Phi^\dagger e^V \Phi$ operators, where $\Phi$ represents 
a generic MSSM chiral superfield, and the additional dimension four operators $Le^VH_d^\dagger$
that violate $R$ parity. In all cases, $V$ should be chosen as the  correct linear combination of 
individual vector superfields to insure gauge invariance. 
 At dimension five level, we have three additional 
structures that are allowed by all gauge symmetries and $R$-parity,
\be
{\cal K}^{(5)} = c_{u}QUH_d^\dagger  + c_{d}QDH_u^\dagger + c_{e}LEH_u^\dagger,
\label{kdim5}
\ee
and several further operators that violate $R$-parity,
\be
{\cal K}^{(5)}_{\not{R}} = c^{\not{R}}_{K1}EH_d H_u^\dagger + c^{\not{R}}_{K2}QUL^\dagger + 
c^{\not{R}}_{K3}UED^\dagger + c^{\not{R}}_{K4}QQD^\dagger.
\ee

At this point, it is important to recall that the equations of motion can be utilized within
the effective Lagrangian to remove various redundancies in the full set of higher-dimensional 
operators listed above. We will work with tree-level matching at the $\La$-threshold and thus, if one leaves aside 
corrections from SUSY breaking, all the structures in  
${\cal K}^{(5)}$ can be reduced on the superfield equations of motion and 
absorbed into ${\cal W}^{(4)}$ and ${\cal W}^{(5)}$. Indeed, in the limit of exact SUSY, the 
superfield equation of motion for {\em e.g.} $H_u^\dagger$ reads
\be
\bar {\rm D}^2 H_u^\dagger \propto -\mu H_d + Y_uQU,
\ee
where $\bar{\rm D}$ is the spinorial derivative. Substituting this into the expression for $K^{(5)}$, we observe that
the operator $LEH_u^\dagger$, for example, reduces to a linear combination of the usual Yukawa structure 
with $\Delta Y_e = \mu c_e$ and the dimension-five superpotential term:
\be
\int d^4\theta c_{e}LEH_u^\dagger \propto \int d^2\theta c_{e}LE {\rm \bar D}^2 H_u^\dagger
\propto \int d^2\theta(-c_e\mu LEH_d +c_eY_u QULE).
\label{eom}
\ee
The inclusion of soft SUSY breaking terms in the equation of motion would change this 
analysis only slightly; new soft-breaking structures such as dimension-four four-sfermion 
interactions $\tilde Q \tilde U \tilde L \tilde E$ and new trilinear terms 
such as $\tilde L \tilde E H^\dagger_u$ \cite{H*} would appear.  Since the analysis of 
higher-dimensional soft-breaking terms goes beyond the scope of the present paper, 
we choose to eliminate all K\"ahler higher dimensional terms via the equations of motion and analyze
only the corrections to superpotential. 

Comparing the $H_uLH_uL$-induced neutrino masses to the characteristic 
mass splitting $\sim (0.01-0.1)$~eV observed in neutrino oscillations, we deduce the corresponding range of the
energy scales $\Lambda_\nu$:
\be
(0.01-0.1)~{\rm eV} \sim c_\nu \langle H_u^2\rangle \qquad\Longrightarrow \qquad\Lambda_{\nu} 
\sim c_\nu^{-1} \sim (10^{14} - 10^{16})~{\rm GeV}.
\ee
The actual mass scale of the new states responsible for generating the effective 
term $H_uLH_uL$ can be lower than 
$\Lambda_\nu$. Indeed, in the see-saw scheme $c_\nu = Y_\nu^2M_R^{-1}$,
and the mass of the right-handed neutrinos $M_R$ can be smaller than 
$\Lambda_\nu$ if $Y_\nu$ is small. A considerably smaller energy scale for $M_R$ than 
$10^{14}$ GeV is indeed suggested by SUSY leptogenesis scenarios \cite{leptogen}.

The mediation of proton decay by the $QQQL$ and $DUUE$ operators has been extensively studied 
over more than two decades in the context of SUSY GUT models. Typically, such 
operators are induced by the exchange of a colour-triplet Higgs superfield, and
therefore the operators are proportional to the square of the Yukawa couplings. For this study, 
we will not go into the details of how such terms were generated, and simply deduce the sensitivity to 
$c_{p1}$ and $c_{p2}$.  The absence of proton-decay at the level of $\Gamma^{-1} > 10^{32}$yr  implies a
rather stringent upper bound on the baryon and lepton number violating couplings $c_{p}$,
\be
\Lambda_{p} \sim c^{-1}_p > 10^{24}~{\rm GeV},
\ee
which is well above the scale of quantum gravity, $10^{19}$GeV. 
The discrepancy in the scales $\Lambda_{p}$ and $\Lambda_\nu$ is 
somewhat problematic for SUSY GUTs, and is part of the  doublet-triplet 
splitting problem.  In any event, the disparity 
between $\Lambda_{p}$ and $\Lambda_\nu$ clearly illustrates the fact that the 
energy scales associated with the effective operators in (\ref{dim5}) could be widely 
different, and thus motivates  a dedicated study to determine the sensitivity to 
$c_{qq}$, $c_{qe}$ and $c_h$.

\section{Induced operators at the SUSY threshold}

We begin our analysis by making explicit the colour and flavour structure of 
the new dimension five operators.  It is easy to see that the $SU(2)$ indices can be contracted 
in only one way,  via the antisymmetric tensor $\epsilon_{ij}$. 
Therefore, we  suppress these indices in the expression below:
\ba
{\cal W} &=& {\cal W}_{\rm MSSM} + \fr{y_h}{\Lambda_{h}}H_dH_uH_dH_u +
\fr{Y^{qe}_{ijkl}}{\Lambda_{qe}}(U_i Q_j )E_k L_l  \nonumber\\
 &&\;\;\;\;\;\;+ 
\fr{Y^{qq}_{ijkl}}{\Lambda_{qq}}(U_iQ_{j}) (D_k Q_{l} )+
\fr{\tilde Y^{qq}_{ijkl}}{\Lambda_{qq}}(U_it^AQ_{j}) (D_kt^AQ_{l}).\label{qule}
\ea
Here $y_h$, $Y_{qe}$, $Y_{qq}$ and $\tilde Y_{qq}$ are dimensionless coefficients
with the latter three also being tensors in flavour space, while the $\Lambda$'s are the 
corresponding energy scales. The parentheses $(...)$ in (\ref{qule}) denote the contraction of 
colour indices. Note, that for the case of one
generation there is only one way of arranging the $SU(3)$ indices, 
as $(Qt^AU) (Qt^A D)$ reduces to $(QU) (Q D)$ upon the 
use of the completeness relation for the generators of $SU(3)$. 

From the superfield formulation of Eq. (\ref{qule}), 
one can easily move to the component form using the standard rules 
of supersymmetric field theories. However, the full interaction 
Lagrangian resulting from (\ref{qule}) is quite cumbersome, and we 
will quote only those terms that are $\sim 1/\Lambda$ and of potential phenomenological 
importance, namely the terms in the Lagrangian that involve two SM fermions and 
two sfermions. As an example, the $QULE$ operator in the superpotential generates the following 
semi-leptonic two fermion - two boson interaction terms:
\ba
\int d^2 \theta \, QULE \supset \bar U Q \tilde L \tilde E^* - \bar E L \tilde Q \tilde U^*+
\bar U E^c \tilde Q \tilde L - \bar U L \tilde Q \tilde E^* 
+\bar Q^cL \tilde U^* \tilde E^* + \bar E Q \tilde L \tilde U^*.
\label{2q2l}
\ea
In this expression, letters with a tilde atop denote sfermions, and four-dimensional spinors 
are used for fermions with $Q$ being the left-handed quark doublet and $Q^c$ 
its charge conjugate, etc. The generalization of (\ref{2q2l}) to the rest of the 
operators in (\ref{qule}) is straightforward. 

At the next step, we integrate out the squarks and sleptons  to obtain  operators 
composed from SM fields alone, or to be more precise, from the fields of a type II 
two-Higgs doublet model. This procedure is facilitated by the observation that the first two terms in 
(\ref{2q2l}) have a close resemblance to the LR squark and slepton mixing terms, with the only 
difference being that instead of the usual
$m_{e(u)}\mu \tan^{\pm1}\beta$ and/or $A_{u(e)}m_{u_e}$ mixing coefficients one has  dimension three 
fermion bilinear insertions $\bar U Q$ and $\bar E L$. It is then clear that $\tilde L \tilde E^*$
and $\tilde Q \tilde U^*$ can be integrated out straightforwardly encountering 
loop integrals that are common in the MSSM literature. 
Notice that only in the first two terms in (\ref{2q2l}) can the sfermions be integrated out at one 
loop, as the remaining terms contain a slepton and a squark, and so integrating them out 
requires at least two loops.

\subsection{Corrections to the SM fermion masses}

\begin{figure}
\centerline{\includegraphics[width=6cm]{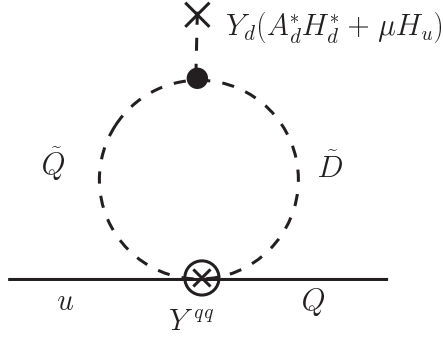}}
 \caption{\footnotesize A one-loop correction to the  masses of SM fermions generated 
by the dimension-5 operators in the superpotential. Here and below crossed vertices stand for the two-fermion--two-boson 
operators generated by the dimension five operators.}
\label{f1} 
\end{figure}

The SM operators of lowest dimension that are of phenomenological interest are the 
fermion masses. In Figure~\ref{f1}, we show the one-loop diagrams that lead 
to the logarithmic renormalization of the fermion masses. Cutting the 
ultraviolet divergence at the corresponding threshold $\Lambda$, we arrive at the 
following expression for fermion masses corrected by the dimension five operators:
\begin{eqnarray}
\label{delta_m}
(M_e)_{ij} &=& (M_e^{(0)})_{ij} + Y^{qe}_{klij}(M_u^{(0)})^*_{kl}
~\fr{3\ln(\Lambda_{qe}/m_{\rm sq})}{8\pi^2 \Lambda_{qe}}
(A_u^* + \mu \cot\beta)\\
(M_d)_{ij} &=& (M_d^{(0)})_{ij} +K^{qq}_{klij}(M_u^{(0)})^*_{kl}
~\fr{\ln(\Lambda_{qq}/m_{\rm sq})}{4\pi^2 \Lambda_{qq}}
(A_u^* + \mu \cot\beta)\nonumber\\
(M_u)_{ij} &=& (M_u^{(0)})_{ij} + Y^{qe}_{ijkl}(M_e^{(0)})_{kl}
~\fr{\ln(\Lambda_{qe}/m_{\rm sl})}{8\pi^2 \Lambda_{qe}}
(A_e + \mu \tan\beta)\nonumber\\
&&+K^{qq}_{ijkl}(M_u^{(0)})^*_{kl}
~\fr{\ln(\Lambda_{qq}/m_{\rm sq})}{4\pi^2 \Lambda_{qq}}(A_u^* + \mu \cot\beta)
\nonumber
\end{eqnarray} 
with implicit summation over the repeated flavour indices, and we have also defined the
combination,
\be
 K^{qq} \equiv  Y^{qq} - \frac{2}{3}\tilde{Y}^{qq},
\ee
that will reappear again below. $M^{(0)}_{e,d,u}$ denote the unperturbed 
mass matrices arising from dimension four terms in the superpotential.

Some of the mass corrections in (\ref{delta_m}) correspond to new ``non-holomorphic" operators
such as $\bar U Q H_d^\dagger$, which break supersymmetry, and 
scale as $\Delta m/m \sim (A/\Lambda) \times \log \Lambda$. 
The other set of corrections survive in the limit of unbroken SUSY, 
scaling as $\Delta m/m \sim (\mu/\Lambda) \times \log \Lambda$.
This is a correction to the standard mass term in the superpotential, $UQH_u$
generated by the dimension five operator. Given the non-renormalization 
theorem for the superpotential  \cite{WB}, it may look surprising that 
such corrections could arise at all. A more careful look at the diagram in Fig.~\ref{f1} 
reveals that  it is the dimension five K\"ahler term $QUH_d^\dagger$ that receives a 
logarithmic loop correction, leading to the quark mass correction in (\ref{delta_m}) 
upon the use of the equation of motion (\ref{eom}).

\subsection{Dipole operators}

\begin{figure}
\centerline{\includegraphics[width=6cm]{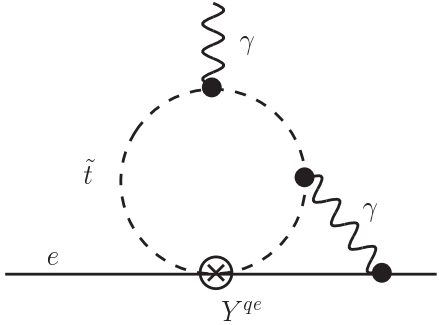}}
 \caption{\footnotesize A representative of the two-loop SUSY threshold diagrams that generate 
dipole amplitudes and contribute to EDMs, $\mu\to e\gamma$, the anomalous magnetic moment 
of the muon, etc. }
\label{f2} 
\end{figure}

Dimension five dipole operators first arise at two-loop order via integrating out the heavy-flavour 
squarks from $\bar E L \tilde U^* \tilde Q$, as in Fig.~\ref{f2}. The results for these diagrams can be deduced from the
calculations of the two-loop Barr-Zee-type supersymmetric diagrams in the limit of large 
pseudoscalar mass \cite{CKP}. In the charged lepton sector they result in 
\be
\label{dipole}
{\cal L}_{e} =    
\fr{A_u +\mu\cot\beta}{\Lambda^{qe}m_{\rm sq}^2} \fr{e\alpha}{12\pi^3} 
(M_u)^*_{kl}Y^{qe}_{klij}\bar E_i (F\sigma) P_L E_j +(h.c.),
\ee
where we treated $LR$ squark mixing as a mass insertion, and used $P_L = \fr{1-\gamma_5}{2}$ and 
$(F\sigma) = F_{\mu\nu}\sigma^{\mu\nu}$.  In the quark sector the corresponding results are more cumbersome 
to write down due to a large number of possible diagrams.

\subsection{Semileptonic four-fermion operators}

\begin{figure}
\centerline{\includegraphics[width=12cm]{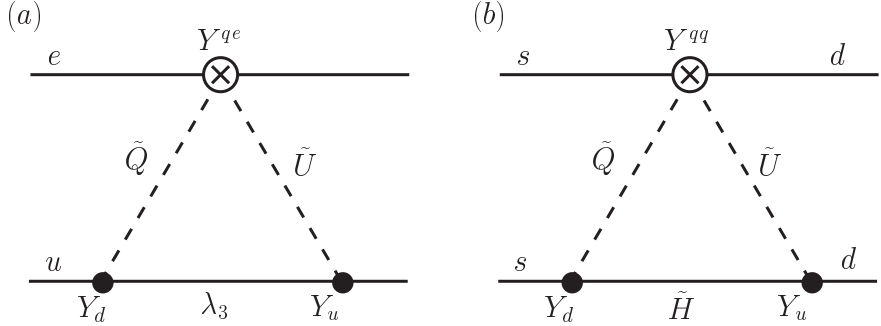}}
 \caption{\footnotesize One-loop SUSY threshold diagrams that generate 
dimension six four-fermion operators composed from the SM fields. Diagram (a) is a 
squark-gluino loop giving rise to a semi-leptonic operator, and diagram (b) is a 
squark-Higgsino loop leading to a four-fermion operator in the down-quark sector.  }
\label{f3} 
\end{figure}

Going up another dimension, we now consider dimension-six four-fermion 
operators composed from the SM fermion fields 
generated by the operators (\ref{qule}). Two representatives 
of the relevant one-loop diagrams are shown in Figure~\ref{f3}. 
The loop functions entering these calculations are identical to those found 
in the calculation of the corrections to the SM fermion masses arising from the
SUSY threshold \cite{masscorr}. We will generalize the results of \cite{prs} by
working with the full loop function \cite{masscorr}, 
\be
 I(x,y,z) = - \frac{xy\ln(x/y) + yz\ln(y/z) + zx\ln(z/x)}{(x-y)(y-z)(z-x)},
\ee
which satisfies
\be
   I(z,z,z) =  \frac{1}{2z},
\ee
allowing us to consider several benchmark SUSY spectra later on.
All the SUSY masses, $m_{\rm sq}$, $m_{\rm sl}$, $M_i$ 
and the $\mu$ parameter are considered to be
somewhat larger than $M_W$, so that the effects of gaugino-Higgsino mixing in the chargino 
and neutralino sector are not particularly important for the values of the 
loop integrals.

Integrating out gauginos and sfermions at one-loop level, we find the 
following semileptonic operators, sourced by the $QULE$ term in the superpotential,
\be
{\cal L}_{qe} = \fr{1}{\Lambda_{qe}}\left[\fr{2\alpha_s}{3\pi} 
M_3^* I(m_{\tilde{u}_1}^2,m_{\tilde{u}_2}^2,|M_3|^2) -
\fr{\alpha_1}{4\pi} 
M_1^* I(m_{\tilde{e}_1}^2,m_{\tilde{e}_2}^2, |M_1|^2) \right]
Y^{qe}_{ijkl}\bar U_i Q_j \bar E_k L_l   + (h.c.).
\label{qqll}
\ee
In this expression, we retained the gluino-squark contribution as the largest in the 
squark sector and the sfermion-bino contribution in the lepton sector. If all SUSY masses
are approximately the same, then the second term in the square bracket of Eq.~(\ref{qqll})
is subdominant, but this may not be the case if the masses in the slepton-bino sector 
are significantly lighter than in the squark-gluino sector.
Notice the absence of contributions from $SU(2)$ gauginos, that turn out to 
be suppressed by additional power(s) of $M_W/m_{\rm soft}$. 
Finally, as expected the overall coefficient in front of
the semileptonic operator (\ref{qqll}) scales as $(\Lambda m_{\rm soft})^{-1}$. 

\subsection{Four-quark operators}

Purely hadronic operators in (\ref{qule}) give rise to the following four-quark 
effective operators upon integrating out gluinos and squarks:
\begin{eqnarray}
{\cal L}_{qq} &=& \fr{1}{\Lambda_{qq}}\fr{\alpha_s}{6\pi} 
M_3^* I(m_{\tilde{q}_1}^2,m_{\tilde{q}_2}^2,|M_3|^2)\nonumber \\
&& \times K^{qq}
\left(\fr{8}{3}(\bar U Q) (\bar D Q)+
(\bar U t^A Q) (\bar D t^A Q)\right)
 + (h.c.),
\label{qqqq}
\end{eqnarray}
where  the summation over flavour is carried out exactly as in Eq.~(\ref{qule}). 

It is well-known that the strongest constraints on FCNC in the quark sector often arise 
from $\Delta F = 2$ amplitudes in the down-squark sector that contribute to the 
mixing of neutral $K$ and $B$ mesons. It is easy to see that such amplitudes are not present in 
Eq. (\ref{qqqq}) where two of the quarks are always of the up-type. Of course, they can be 
converted to down-type quarks at the expense of an additional loop with $W$-bosons, but
this introduces an additional numerical suppression. In any event, the conversion of the right-handed 
quark field $U$ into a $D$ field would necessarily require additional Yukawa suppression. 
There is, however, a more-direct one-loop SUSY threshold diagram that can give rise to 
$\Delta F =2$ amplitudes in the down-quark sector.
As shown in Figure~\ref{f3}b, it consists of a Higgsino--up-squark loop. The result 
for this diagram,
\begin{eqnarray}\label{dddd}
{\cal L}_{dd} &=& \fr{1}{\Lambda_{qe}}\fr{1}{8\pi^2} 
\mu^* I(m_{\tilde{u}_1}^2,m_{\tilde{u}_2}^2,|\mu|^2)(Y^*_u)_{im}(Y^*_d)_{nj}
\nonumber\\
&&\times  K^{qq}_{ijkl}\left[\fr{1}{3}(\bar Q_m D_n) (\bar D_k Q_l)
- (\bar Q_m t^A D_n) (\bar D_k t^A Q_l)\right]
  + (h.c.),
\end{eqnarray}
inevitably contains additional suppression by the Yukawa couplings of the up and down 
type quarks originating from the Higgsino-fermion-sfermion vertices. 

Notice that in the limit $\mu\gg m_{sq}$ the overall coefficient in equation 
(\ref{dddd}) scales as $(\Lambda\mu)^{-1}\log(\mu/m_{sq})$ and thus has only mild 
dependence on the soft-breaking scale. In this case, the operator (\ref{dddd}) must
have an explicit superfield generalization. Indeed, it is easy to see that in 
this limit (\ref{dddd}) corresponds to a dimension six  K\"ahler term:
$Q^\dagger DD^\dagger Q$.

\subsection{Modifications to the Higgs sector and sparticle spectrum}

Thus far, we have not considered the consequences of the presence of the first operator in (\ref{qule}), 
which consists entirely of Higgs superfields. Its most obvious implication is a modification of the Higgs potential and 
the sparticle spectrum. The addition to the Higgs potential, linear in $y_h$, has a simple form:
\be
\Delta V_h = -\fr{2\mu^*y_h}{\Lambda_h}\left[(H_u^\dagger H_u) + (H_d^\dagger H_d)\right](H_dH_u)
+ (h.c.).
\label{vh}
\ee
If $\mu^*y_h$ has a cumulative phase, this would create mixing between 
$A$ and the $h$, $H$ bosons that violate $CP$ symmetry.  However, its most important consequence
for our study here will be an induced complex shift of the bilinear soft parameter $m_{12}^2$, which
enters one-loop contributions for fermion EDMs. 

The mixing of left- and right-handed sfermions is also affected by this term.
In addition to the usual $\mu$ or $A$-proportional mixing, we 
have the following contribution to the mixing matrix element of $\tilde u_L$ and $\tilde u_R$,
\be
\de(M_{\tilde u}^2)_{LR} = - m_u \fr{y_h v_{SM}^2}{\Lambda_h}\cos^2\beta,
\label{mlr}
\ee
and analogous formulae for $\tilde e$ and $\tilde d$ with the $\cos\beta \rightarrow \sin\beta$
substitution. In this expression, $v_{SM}^2 = 4M_W^2/g^2_W$ corresponds to the SM 
Higgs v.e.v. 

The neutralino mass matrix also receives two new (complex) entries, i.e. 
Majorana masses for the neutral components of $\tilde H_u$ and $\tilde H_d$,
proportional to $y_h\lambda_h^{-1}v_{SM}^2\cos^2\beta$ and $y_h\lambda_h^{-1}v_{SM}^2\sin^2\beta$
respectively.

\section{Phenomenological consequences and sensitivity to $\Lambda$}

In this section, we estimate the sensitivity of various experimental searches
to the energy scales $\Lambda^{qe}$ and $\Lambda^{qq}$. Of course, one of the most important 
issues is then the assumed flavour structure of the new couplings $Y^{qe}$, $Y^{qq}$ and $\tilde Y^{qq}$. 
Since we are thinking of $\La$ as an intermediate scale and wish to explore the full reach of precision measurements,
we will make the generous assumption that these coefficients are complex, of order one, and {\em do not factorize}
into products of Yukawa matrices in the superpotential: $Y^{qe} \neq Y_u Y_e$. It is clear that a much more restrictive
assumption, {\it e.g.} minimal flavour violation, would dramatically reduce the sensitivity to these 
operators, but we will not explore this option here.

\subsection{Naturalness bounds -- fermion masses and the $\theta$-term}

With the above assumption on flavour structure, we should first investigate the requirement that the corrections to 
masses of the SM fermions do not exceed their measured values, as otherwise we will face a new fine-tuning 
problem in the flavour sector. 
Taking $(M_u A_u)_{kl} =  (M_u A_u)_{33} \sim m_tA_t \sim 175 \, {\rm GeV}
\times 300 \, {\rm GeV}$
and using the expression for $\Delta m_e$ in (\ref{delta_m}), we arrive at the following estimate,
\be
\Delta m_e \sim \fr{3 m_t A_t}{8\pi^2\Lambda^{qe}}\ln\left(\fr{\La^{qe}}{m_{sq}}\right)
\sim 1 {\rm MeV} \times \fr{10^7\, {\rm GeV}}{\Lambda^{qe}},
\ee
which clearly suggests that the `naturalness' scale for the new physics 
encoded in semileptonic dimension five operators in the superpotential 
is on the order of $10^7$~GeV, while the analogous sensitivity in the 
squark sector is slightly lower, $\Lambda^{qq}\sim 10^6$~GeV. This high naturalness scale is
simply a restatement of the small Yukawa couplings for the light SM fermions. However, perhaps surprisingly,
we will see below that this sensitivity is not the dominant constraint on the threshold scale.

Before we proceed to estimate the effects induced by four-fermion operators, 
we would like to consider the effective shift of the QCD $\theta$-angle due to the mass corrections
(\ref{delta_m}). Assuming an arbitrary overall phase for the $Y^{qq}$ matrices relative to the phases of 
the eigenvalues of $Y_u$ and $Y_d$, one typically finds the 
following shift of the $\bar \theta$ parameter,
\be
\Delta \bar\theta \sim \fr{{\rm Im}(m_d)}{m_d} \sim 
\fr{ {\rm Im}(Y^{qq}m_tA_t)}{4\pi^2 m_d\Lambda^{qq}}  \ln\left(\fr{\La^{qe}}{m_{sq}}\right)
\sim 10^{-10}\times 
\fr{10^{17}~{\rm GeV}}{\Lambda^{qq}}.
\label{delta_th}
\ee
This is a remarkable sensitivity of $\Delta \theta$ to new sources of $CP$ and flavour violation, 
and can translate into a strong bound on $\Lambda^{qq}$ depending on how the 
strong $CP$ problem is addressed. If it is solved by an axion, there are no consequences of 
(\ref{delta_th}). However, in other possible approaches $\bar\theta\simeq 0$ is engineered by hand,
using {\em e.g.} discrete symmetries at high energies \cite{discrete}. In this case, 
dimension five operators can pose a potential threat up to the energies $\Lambda^{qq} \sim 
10^{17}$ GeV, which by itself is very remarkable. Future progress in measuring the electric 
dipole moments (EDMs) of neutrons and heavy atoms \cite{PR2005}, can clearly bring this scale up to the 
Planck scale and beyond.

\subsection{Electric dipole moments from four-fermion operators}

Electric dipole moments (EDMs) of the neutron \cite{n} and heavy 
atoms and molecules \cite{Tl,Hg,TlF,Xe,Cs,YbF,PbO} are primary  
observables in probing for sources of flavour-neutral $CP$ violation. 
The high degree of precision with which various experiments have put 
limits on possible EDMs translates into stringent constraints on a variety of 
extensions of the Standard Model at and above the electroweak scale (see, e.g. \cite{KL,PR2005}). 
Currently, the strongest constraints on $CP$-violating parameters arise 
from the atomic EDMs of thallium \cite{Tl} and mercury \cite{Hg}, and that of 
the neutron \cite{n}:
\ba
|d_{\rm Tl}| &<& 9 \times 10^{-25} e\, {\rm cm} \nonumber\\
|d_{\rm Hg}| &<& 2   \times 10^{-28}  e\, {\rm cm}     \\
|d_n|  &<&   3\times 10^{-26} e\, {\rm cm}.\nonumber
\label{explimit}
\ea

When $\bar\theta$ is removed by an appropriate symmetry, 
the EDMs are mediated by higher-dimensional operators. Both 
(\ref{qqll}) and (\ref{qqqq}) are capable of inducing the 
atomic/nuclear EDMs if the overall coefficients contain an extra 
phase relative to the quark masses. 
Restricting Eq.~(\ref{qule}) to the first generation and dropping the 
$U(1)$ contribution, we find the following 
$CP$-odd operator:
\be
{\cal L}_{CP-odd} = -
\fr{1}{\Lambda_{qe}}\fr{\alpha_s}{3\pi} 
|M_3Y^{ue}|   I(m_{\tilde{u}_1}^2,m_{\tilde{u}_2}^2,|M_3|^2)   \sin\delta
\times \left[(\bar uu) (\bar ei\gamma_5 e) + (\bar ui\gamma_5u)( \bar e e) \right],
\ee
with the $CP$-violating phase $\delta = {\rm Arg}(M_3^*Y^{qe})$ in a 
basis with real $m_e$ and $m_u$. Taking into account the $QCD$ running from the superpartner mass scale 
to $1$~GeV,  and upon the use of hadronic matrix elements over nucleon states,
$\langle N|\bar uu |N\rangle$ and $\langle N|\bar ui \gamma_5u |N\rangle$, 
we can make a connection to the $C_S$ and $C_P$ coefficients in the 
effective $CP$-odd electron-nucleon Lagrangian, 
\be
 {\cal L} = C_S \bar NN \bar e i \gamma_5 e + C_P \bar Ni \gamma_5N \bar e  e.
 \ee
The isospin singlet part of the $C_S$ coefficient is given by
\ba
C_S &=& -\frac{4\al_s}{3\pi \La^{qe}} {\rm Im}(M_3^* Y^{uuee})I(m_{\tilde{u}_1}^2,m_{\tilde{u}_2}^2,|M_3|^2) \nonumber\\
    &\sim& 2\times 10^{-4} (1\,{\rm GeV}\times\Lambda^{qe})^{-1},
\label{CS}
\ea
where in the latter equality we also assumed maximal violation of $CP$,
 $ |Y^{qe}|_{(M_Z)}\sim \sin \delta \sim O(1)$,  and chose the 
 superpartner masses degenerate at 300~{\rm GeV}.
The quark matrix element, $\langle N| (\bar uu + \bar dd)/2|N\rangle \simeq 4$, 
is in accord with standard values for the quark masses and the nucleon $\sigma$-term. 

Using the same assumptions, and the pseudoscalar matrix element
over the neutron,
$\langle n|\bar ui \gamma_5u |n\rangle \simeq -0.4(m_N/m_u)\bar n i \gamma_5 n$,
we obtain a similar expression for the neutron $C_P$ coefficient, 
\ba
C_P &=& \frac{\al_s}{6\pi \La^{qe}} \left(0.4\frac{m_n}{m_u}\right){\rm Im}(M_3^* Y^{uuee})I(m_{\tilde{u}_1}^2,m_{\tilde{u}_2}^2,|M_3|^2) 
 \nonumber\\
&\sim& 4\times 10^{-3} (1\,{\rm GeV}\times\Lambda^{qe})^{-1}.
\label{CP}
\ea

Comparing (\ref{CS}) and (\ref{CP}) to the limits on $C_S$ and $C_P$ deduced from 
the bounds on the EDMs of Tl and Hg \cite{FG}, we obtain the following sensitivity 
to the energy scale $\Lambda^{qe}$,
\begin{eqnarray}
\label{cslimit}
\Lambda^{qe} &\ga& 3 \times 10^8 ~{\rm GeV} ~~~~~~~~~~{\rm from~ Tl~ EDM} \\
\Lambda^{qe} &\ga& 1.5 \times 10^8 ~{\rm GeV} ~~~~~~~~{\rm from~ Hg~ EDM}
\end{eqnarray}
These are remarkably large scales, and indeed not far from the intermediate scale suggested by 
neutrino physics. In fact, the next generation of atomic/molecular EDM experiments have the chance of 
increasing this scale by two-three orders of magnitude which would put it
close to the scales often suggested for right-handed neutrino masses.

Going over to purely hadronic $CP$-violating operators, {\it e.g.} $C_{ud} (\bar d i \gamma_5 d)(\bar u u)$,
we note that these would induce the EDMs of neutrons, and EDMs of diamagnetic atoms 
mediated by the Schiff nuclear moment $S(\bar{g}_{\pi NN})$. In particular, we have for the $CP$-odd 
isovector pion-nucleon coupling,
\be
 \bar{g}^{(1)}_{\pi NN} = - 4 \times 10^{-2} \frac{C_{ud}}{m_d}, 
\ee
with 
\be
C_{ud} = -\frac{\al_s}{9\pi \La^{qq}} {\rm Im}(M_3^* Y^{uudd})I(m_{\tilde{u}_1}^2,m_{\tilde{u}_2}^2,|M_3|^2),
\ee
obtained as for the semileptonic operators above.
The typical sensitivity to $\Lambda^{qq}$ in this case is 
somewhat lower than in the case of semi-leptonic operators,
\be
\Lambda^{qq} ~\ga~ 3\times 10^7 ~{\rm GeV} ~~~~~~~~{\rm from~ Hg~ EDM}\nonumber.
\label{edmqq}
\ee

Semileptonic operators involving heavy quark superfields are also tightly constrained 
by experiment, via the two-loop diagrams of Fig.~\ref{f2}. Assuming no additional $CP$ violation 
in the soft-breaking sector and taking into account only 
the stop loops, we obtain the following result for the EDM of the electron:
\be
d_e = e\fr{\alpha}{12\pi^3}~\fr{{\rm Im}(Y^{ttee})}{\Lambda^{qe}}\frac{m_t|A_t - \mu^*\cot\beta|}{|m_{\tilde{t}_1}^2 - m_{\tilde{t}_2}^2|}
\ln\left(\fr{m_{\tilde{t}_1}^2}{m_{\tilde{t}_2}^2}\right),
\label{de2l}
\ee
where $m_{\tilde{t}_1}$ and $m_{\tilde{t}_2}$ are the stop masses in the physical basis.
Assuming maximal $CP$-odd phases and large stop mixing, we arrive at the following estimate for $d_e$,
\be
d_e \sim  \fr{10^8\,{\rm GeV}}{\Lambda^{qe}}\times 10^{-27} ~e\,{\rm cm},
\ee
which together with the sensitivity to the electron EDM inferred from the constraint on
$d_{\rm Tl}$, $|d_e|\la 1.6\times 10^{-27} e\, {\rm cm} $, translates to 
$$
\Lambda^{qe}\ga 6 \times 10^7\,{\rm GeV}.
$$
Expressions similar to (\ref{de2l}) can be obtained for the quark EDMs and color
EDMs, furnishing similar sensitivity to $\Lambda^{qq}$.

\subsection{Lepton flavour violation}

Searches for lepton-flavour violation, such as the decay $\mu\to e\gamma$ and 
$\mu\to e$ conversion on nuclei have resulted in stringent upper bounds on the 
corresponding branching ratio \cite{muegamma} and the rate of conversion 
normalized on capture rate \cite{sindrum}\footnote{A recent announcement from the SINDRUM II collaboration
suggests a slightly stronger constraint, $R({\mu \to e^-~  {\rm on ~Au}}) < 7\times 10^{-13}$ \cite{sindrum2}.}:
\begin{eqnarray}
&& {\rm Br}({\mu\to e\gamma}) < 1.2\times 10^{-11} \\
&&R({\mu \to e^-~  {\rm on ~Ti}}) < 4.3\times 10^{-12} .
\end{eqnarray}

Focussing first on $\mu\to e$ conversion, one can deduce the sensitivity of these searches
to the energy scale of the semileptonic  operators (\ref{qule}) as the conversion is mediated by the 
$(\bar uu)(\bar e i \gamma_5 \mu)$ and $(\bar uu) (\bar e \mu)$ operators, and thus 
involves the same matrix elements as does $C_S$. Indeed, the characteristic amplitude for the scalar operator
has the form
\be
 \frac{G_F}{\sqrt{2}} \et_{e\mu} = -\frac{4\al_s}{3\pi \La^{qe}} {\rm Im}(M_3^* Y^{uue\mu})I(m_{\tilde{u}_1}^2,m_{\tilde{u}_2}^2,|M_3|^2).
 \ee
Using the bounds on such scalar operators
derived elsewhere (see {\em e.g.} \cite{faessler}),
we conclude that $\mu\to e$ conversion currently probes energy scales as high as 
\be
\Lambda^{qe} \ga 1\times 10^8 \,{\rm GeV}~~~~~{\rm from}~{\mu^- \to e^-~  {\rm on ~Ti}}.
\label{muelimit}
\ee
However, it is important to note that the bound on the conversion rate is necessarily proportional to $(\et_{e\mu})^2$ and thus
these effects decouple as $(\La^{qe})^{-2}$ in contrast to the linear decoupling of the EDMs.

A sensitivity to slightly lower scales arises from the two-loop--mediated $\mu\to e\gamma$ 
process. We have 
\be
 {\rm Br}({\mu\to e\gamma}) = 384\,\pi^2 \frac{\mu_{e\mu}^2}{4G_F^2 m_{\mu}^2},
 \ee
 with the transition amplitude generated in the same manner as $d_e$,
 \be
  \mu_{e\mu} = \fr{\alpha}{12\pi^3}~\fr{{\rm Re}(Y^{ttee})}{\Lambda^{qe}}\frac{m_t|A_t - \mu^*\cot\beta|}{|m_{\tilde{t}_1}^2 - m_{\tilde{t}_2}^2|}
\ln\left(\fr{m_{\tilde{t}_1}^2}{m_{\tilde{t}_2}^2}\right), 
\ee
where once again the sensitivity in the branching fraction 
is weakened relative to the EDMs by quadratic decoupling with the threshold scale.

Future progress in lepton flavour violation searches should be able to extend the reach of 
these probes by one-two orders of magnitude. Disregarding a factor of a few between (\ref{cslimit})
and (\ref{muelimit}), we conclude that currently the EDMs and searches for lepton flavour
violation probe these extensions of the MSSM up to similar energy scales 
of $\sim10^8$ GeV. 

It is also worth noting that sensitivity to $\La^{qe}$, that is somewhat more robust to changing assumptions on 
the flavour structure of  $Y^{ue}$, can also be achieved through comparison of the two modes of charged pion decay 
into first and second generation leptons. The typical sensitivity to 
$\Lambda^{qe}$ in this case could be as large as 
$$
\Lambda^{qe} ~\ga~ 10^4 ~{\rm GeV}~~~~~~~~{\rm from~ \mu-}e 
{\rm ~universality~in~\pi^\pm ~decay }\nonumber.
$$

Finally, the two-loop amplitudes in Fig.~3 would also give corrections to the anomalous magnetic moments 
of $e$ and $\mu$. In the latter case, one can estimate the sensitivity to $\Lambda^{qe}$
as no higher than about 1 TeV.

\subsection{$K$ and $B$ meson mass-difference}

Often, the most constraining piece of experimental information comes from the 
contributions of new physics to the mixing of neutral mesons, $K$ and $B$. 
In the case of generic couplings $Y^{qq}$ and $\tilde Y^{qq}$, the four-fermion 
operators (\ref{dddd}) will contain $(\bar s_R d_L)(\bar s_L d_R)$  and 
$(\bar b_R d_L)(\bar b_L d_R)$ terms. Using a simple vacuum factorization ansatz, we find
\ba
\langle K^0|(\bar d_R s_L)(\bar d_L s_R)|\bar K^0\rangle 
&=& \left[\fr{1}{24}
+\fr 14\left(\fr{m_K}{m_s+m_d}\right)^2\right]m_K f_K^2
\simeq 2 m_K f_K^2,
\nonumber\\
\langle K^0|(\bar d_R t^As_L)(\bar d_L t^As_R)|\bar K^0\rangle 
&=& \fr{1}{18}
m_K f_K^2 =0.055 m_K f_K^2,
\ea
and therefore can neglect $(\bar d_R t^As_L)(\bar d_L t^As_R) $
due to its small matrix element. 
Taking into account the one-loop QCD evolution of 
the $(\bar d_R s_L)(\bar d_L s_R) $ operators from the SUSY threshold scale down to 1 GeV,
\ba
(\bar d_R s_L)(\bar d_L s_R)_{1~{\rm GeV}} \simeq 5 (\bar d_R s_L)(\bar d_L s_R)_{M_Z},
\ea
we can estimate the contribution of additional four-fermion operators to the mass splitting 
of $K$ mesons:
\ba
\label{DmK}
\Delta m_K = -2{\rm Re}\langle K^0|{\cal L}_{dd}|\bar K^0\rangle 
\simeq - 5\fr{m_K f_K^2\mu I(m_{\tilde{u}_1}^2,m_{\tilde{u}_2}^2,|\mu|^2) }{6\pi^2\Lambda^{qq}}
~{\cal Y}_{ds},
\ea
where we took $\mu$ to be real, and introduced the following notation for the relevant combination 
of Yukawa couplings:
\ba
{\cal Y}_{ds}={\rm Re}
(Y^*_u)_{i1}(Y^*_d)_{2j}      K^{qq}_{ij12}
+{\rm Re}(Y_u)_{i2}(Y_d)_{1j}  K^{qq*}_{ij21}.
\label{yds}
\ea
 It is easy to see that the presence of 
$Y_u$ and $Y_d$ in (\ref{yds}) results in a significant numerical suppression of 
the corresponding Wilson coefficients even with $Y^{qq}\sim O(1)$. The minimal suppression
is realized with an intermediate stop-loop, in which case 
the first term in (\ref{yds}) becomes of order $ V_{td}y_t y_s$, while the second 
term is $\sim V_{ts}y_t y_d$. Numerically, this corresponds to
a suppression factor,
\be
{\cal Y}_{ds}\sim 3\times 10^{-4} \times \fr{\tan\beta}{50},
\label{ydsnumber}
\ee
which is also $\tan\beta$-dependent. 

Putting all the factors together, with the same assumption of degenerate soft mass parameters at
300{\rm GeV}, 
we come to a disappointingly weak result:
\be
\Delta m_K \sim 3\times 10^{-6} \, {\rm eV}\times \fr{\tan\beta}{50}\times
\fr{200\,{\rm GeV}}{\Lambda^{qq}},
\label{dmklimit}
\ee
with the actual measured value of the mass splitting being $3.5\times 10^{-6}$eV. 
The calculation of $\Delta m_B$ results in a similar sensitivity level, 
prompting the conclusion that neither $\Delta m_K$ nor $\Delta m_B$ can probe 
the flavour structure of additional dimension five operators beyond the SUSY threshold. 
The $CP$-violating observable $\epsilon_K$ will obviously be more sensitive by almost 
three orders of magnitude, resulting in
\be
\Lambda^{qq} \ga 10^5\,{\rm GeV}\times \fr{\tan\beta}{50}~~~~~~~{\rm  from}~\epsilon_K,
\label{epsk}
\ee
which is still clearly inferior to the sensitivity of EDMs and lepton flavour violation. 
Moreover, it is easy to see that if the complete theory at scales 
$\Lambda$ also provides new dimension-six operators in the K\"ahler potential, the 
possible consequences of those for $\De m_K$ and $\De m_B$  would be considerably more serious that of the
dimension-five operators. We give an explicit example of this in the discussion section.

\subsection{Constraining the  Higgs operator}

The strength of the constraints on $QULE$ and $QUQD$ 
comes primarily from the fact that such operators flip the chirality of 
light fermions without paying the usual price of small Yukawa couplings.
This was a consequence of our assumption on the arbitrary flavour structure
of the dimension-five operators.
Should all transitions from $u_R$ to $u_L$ and $e_R$ to $e_L$ 
be suppressed by $m_f/v$, the constraints (\ref{cslimit}), (\ref{muelimit})
and (\ref{epsk}) would be relaxed all the way to the weak scale and below.
Therefore, it would come as no surprise if the effective operator in the
Higgs potential were to have very weak implications for $CP$-violating physics
and no consequences at all for the flavour-changing transitions. 

We note first of all that the mixing of the left- and right-handed $d$-squarks is 
affected by the Higgs operator (\ref{mlr}).  This feeds into the one-loop $d$-quark EDM diagram, 
where this parameter behaves similarly to the insertion of the complex $A$-term,
$A_d^{eff} \sim y_h v_{SM}^2 \Lambda_h^{-1}$ and leads to a contribution to $d_d$ that does not 
grow with $\tan \beta$.  The typical sensitivity of the neutron and 
mercury EDMs to the imaginary part of the $A_d$ parameter \cite{ourlateststuff}, 
with the superpartner masses in the ballpark of  $\sim 300$ GeV, then implies 
\be
\Lambda_h \ga 1~{\rm TeV}~~~~~~~{\rm maximal~}CP~{\rm violation,~ neutron~EDM}.
\ee
Of course, a mere increase of the superpartner masses to around 1 TeV would 
completely erase this sensitivity.
 
Another possibly interesting $CP$-violating effect would come from the 
admixture of the pseudoscalar Higgs $A$ to the scalars $h$ and $H$  at tree level. 
Subsequent Higgs exchange would then induce the $C_S$ operator \cite{Barr,LP},
or contribute to the two-loop EDM of quarks and electrons \cite{BZ}. 
The latter results in contributions to observable  EDMs that are $\tan\beta$-dependent and 
furnish a sensitivity to $\Lambda_h$ up to a few TeV.

These are relatively minor effects. However, it turns out that significant sensitivity
to this operator can indeed arise through its shift of the Higgs potential (\ref{vh}), and
more specifically the effective shift of the  $m_{12}^2$ parameter,
\be
m_{12}^2 H_u H_d \to \left(m_{12}^2 + \fr{\mu y_hv_{SM}^2}{\Lambda_h}\right)H_u H_d ,
\label{meff}
\ee
assuming the reality of $\mu$. The quantity in the parentheses is
an effective $m_{12}^2$ parameter, which is complex 
on account of Im$(y_h)$. Moreover, its complex phase is enhanced in the large
$\tan\beta$ limit because $m_{12}^2 \simeq m_A^2\tan^{-1}\beta$. The resulting 
phase affects the one-loop SUSY EDM diagrams. Assuming for simplicity a common mass scale 
$m_{\rm soft}$ for sleptons, gauginos, and $\mu$, we have \cite{ourlateststuff}:
\ba
d_e &\sim& \fr{em_e \tan\beta}{16\pi^2m_{\rm soft}^2}\left (\fr{5 g_2^2}{24}+\fr{g_1^2}{24}\right)
 \sin \left[{\rm Arg}\frac{\mu M_2}{ (m_{12}^2)_{\rm eff})}\right] .
\ea
Expanding to leading order in $1/\La_h$, and imposing the present limit on $d_e$, we find
the sensitivity,
\be
\Lambda_h \ga 2\times 10^7 ~{\rm  GeV} \left(\fr{\tan\beta}{50}\right)^2
\left(\fr{300{\rm GeV}}{M_{\rm SUSY}}\right)\left(\fr{300{\rm GeV}}{m_A}\right)^2,
\ee
which reaches impressively high scales for maximal $\tan\beta$.

\section{Constraints within MSSM benchmark scenarios}

A summary of the characteristic sensitivity to the threshold scale $\La$ in different channels is given in Table~2,
assuming generic and degenerate SUSY spectra. 
In this section, we will go somewhat further and examine the variation in sensitivity among a few benchmark SUSY scenarios, with different
spectra chosen in order to satisfy other constraints, including requiring the correct relic LSP density to form dark
matter. We have also included the leading one-loop SUSY evolution of the dimension-five operators from the $\La$-scale
to the soft threshold. These effects are generally on the order of 20-40\%, and the relevant RG equations are included 
in the Appendix.

\begin{table}
\begin{center}
\begin{tabular}{||c|c|c||}
\hline
operator & sensitivity to $\La$ (GeV) & source \\ \hline\hline
$Y^{qe}_{3311}$ & $\sim 10^7$ & naturalness of $m_e$\\
Im($Y^{qq}_{3311})$ & $\sim 10^{17}$ & naturalness of $\bar \theta$, $d_n$\\
Im($Y^{qe}_{ii11}$) & $10^7-10^9$ & Tl, Hg EDMs \\ 
$Y^{qe}_{1112}$, $Y^{qe}_{1121}$ & $10^7-10^8$& $\mu\rightarrow e$ conversion \\ 
Im($Y^{qq}$) & $10^7-10^8$ & Hg EDM \\ 
Im($y_h$) & $10^3-10^8$ & $d_e$ from Tl EDM \\ \hline
\end{tabular}
\end{center}
\caption{Sensitivity to the threshold scale. Note that the naturalness bound on Im$(Y^{qq})$ 
doesn't apply to the axionic solution of the strong $CP$ problem, the best sensitivity to Im$(y_h)$ 
is achieved at maximal $\tan\beta$, and the Hg EDM constraint on Im$(Y^{qq})$ applies when at least one pair 
of quarks belongs to the 1$^{\rm st}$ generation.
}
\label{table1}
\end{table}

The benchmark spectra we have chosen are the following representative SPS points~\cite{sps}:
\begin{itemize}
\item SPS1a -- msugra 
\item SPS2 --  focus point 
\item SPS4 -- large $\tan\beta$ in the funnel region
\item SPS8 -- gauge mediation
\end{itemize}
\begin{figure}
\begin{picture}(450,200)
\put(0,0){\centerline{\includegraphics[width=9cm]{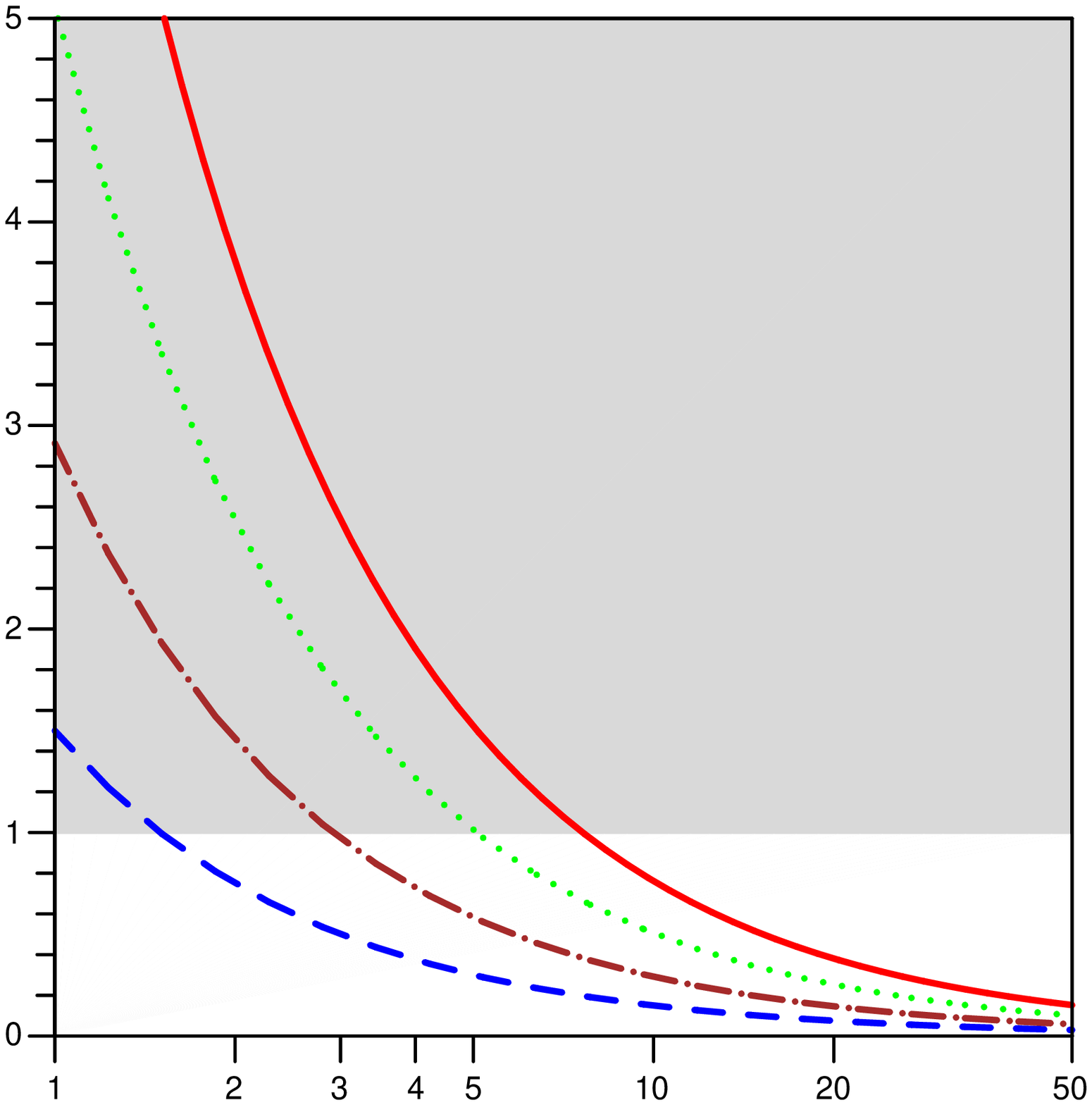}\includegraphics[width=9cm]{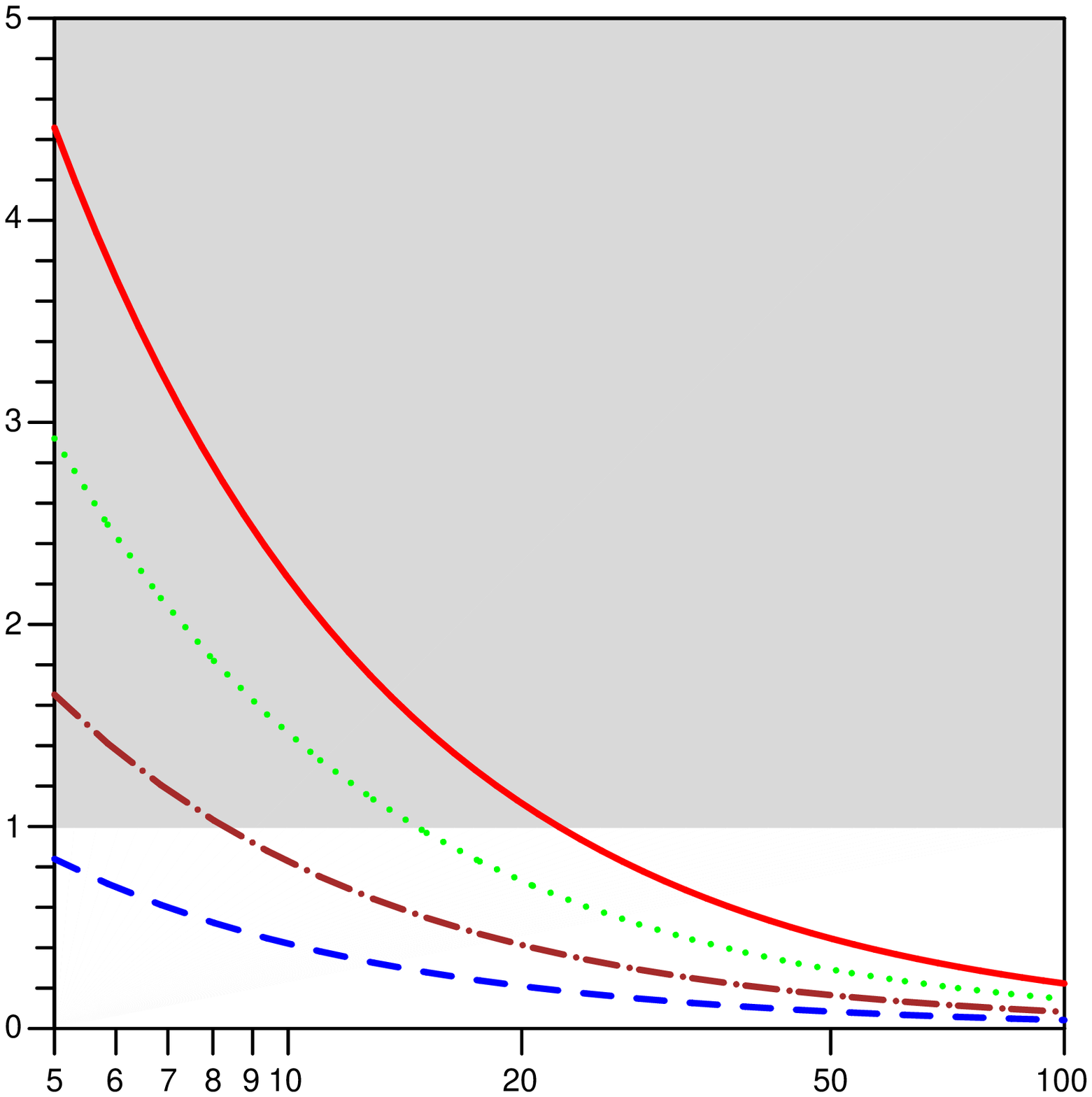}}}
\put(143,0){{$\Lambda$ [$10^7\,$ GeV]}}
\put(0,240){{$d_{Tl}(C_S)$}}
\put(400,0){{$\Lambda$ [$10^5\,$ GeV]}}
\put(258,240){{$d_{Hg}(\bar{g}_{\pi NN})$}}
\end{picture}
 \caption{\footnotesize Constraints on the benchmark scenarios from contributons to $d_{Tl}(C_s)$ and $d_{Hg}(\bar{g})$. In these plots and
 those below, SPS1a = red (solid), SPS2 = blue (dashed), SPS4 = green (dotted), SPS8 = brown (dot-dashed). The shaded region is above the
 current experimental bound.}
\label{b1} 
\end{figure}

\begin{figure}
\begin{picture}(450,200)
\put(0,0){\centerline{\includegraphics[width=9cm]{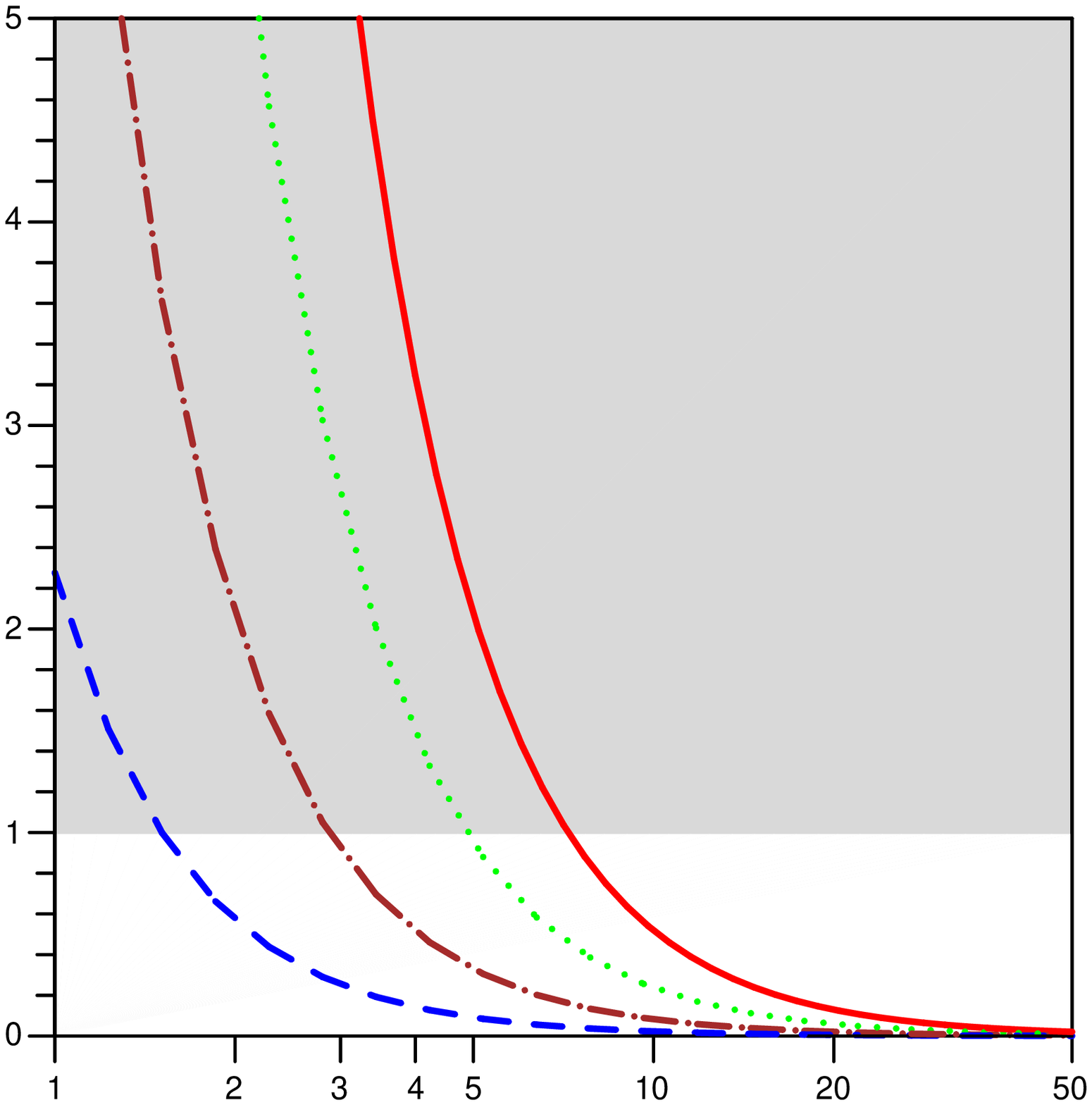}\includegraphics[width=9cm]{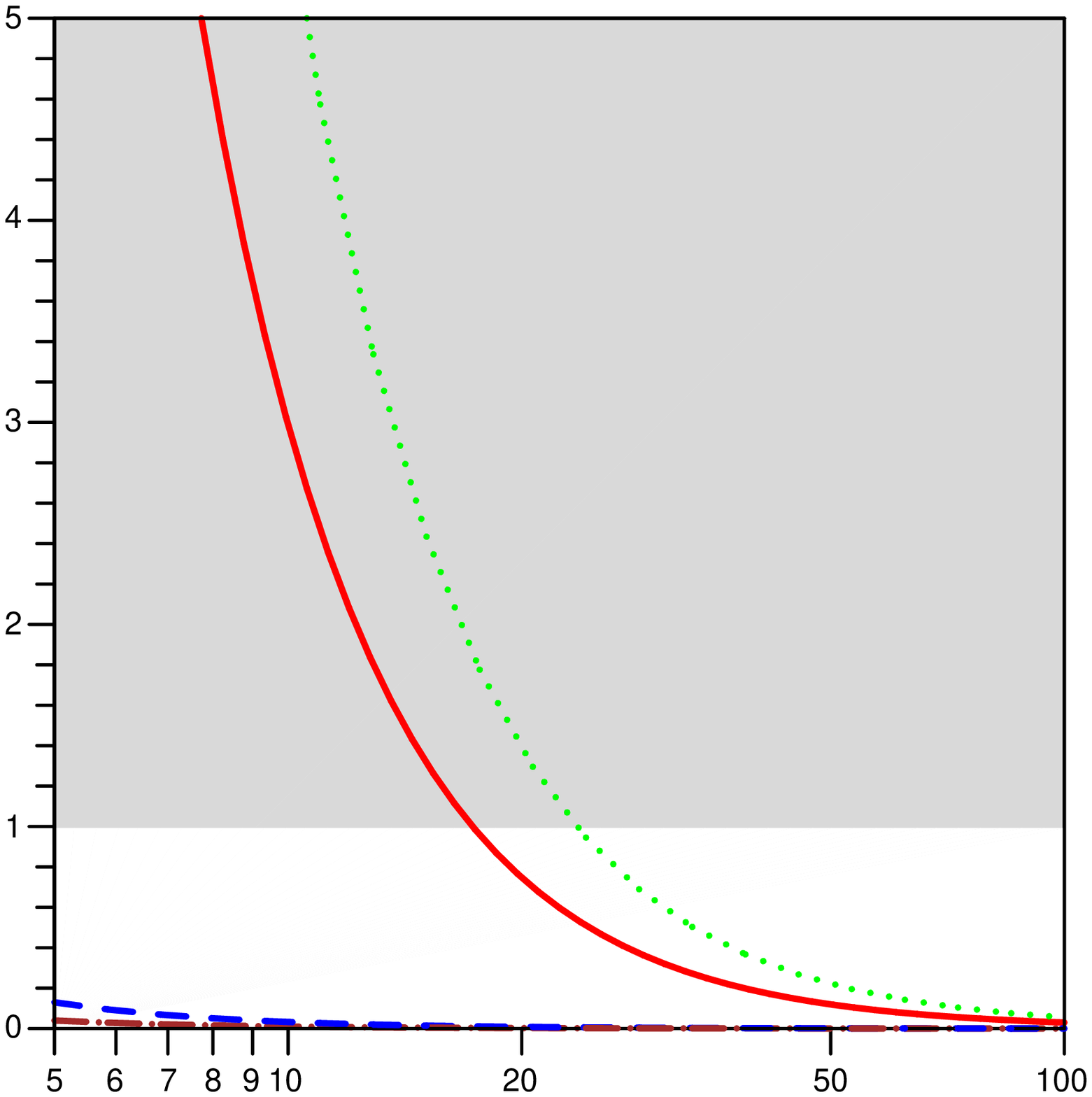}}}
\put(143,0){{$\Lambda$ [$10^7\,$ GeV]}}
\put(0,240){{$R(\mu\rightarrow e)$}}
\put(400,0){{$\Lambda$ [$10^5\,$ GeV]}}
\put(258,240){{$Br(\mu \rightarrow e\gamma)$}}
\end{picture}
 \caption{\footnotesize Constraints on the benchmark scenarios from contributons to  $\mu\rightarrow e$
 conversion and $\mu\rightarrow e \gamma$.}
\label{b2} 
\end{figure}

\begin{figure}
\begin{picture}(450,220)
\put(0,0){\centerline{\includegraphics[width=9cm]{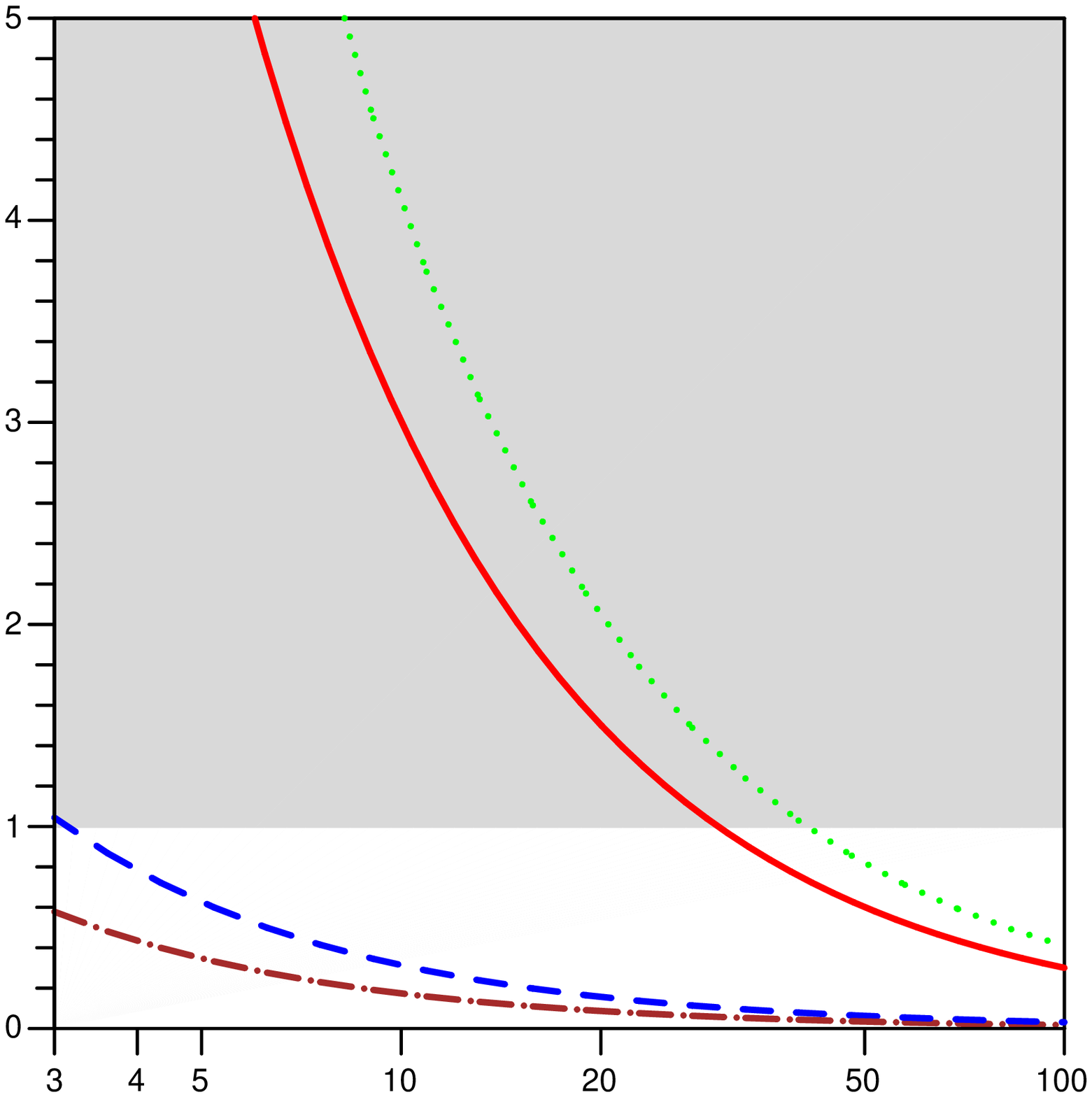}\includegraphics[width=9cm]{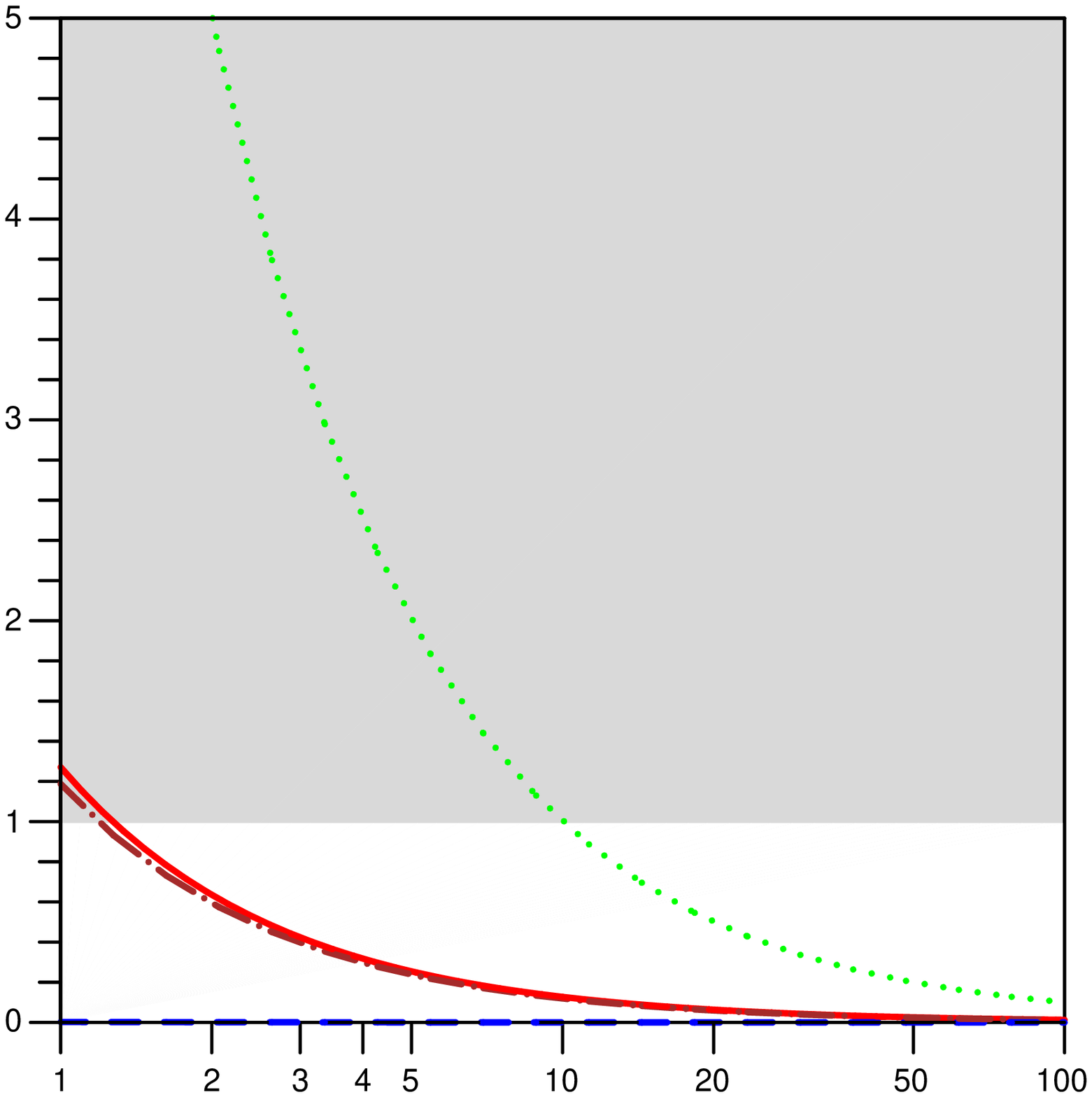}}}
\put(143,0){{$\Lambda$ [$10^7\,$ GeV]}}
\put(0,240){{$d_{Tl}(d_e({\rm 2l}))$}}
\put(400,0){{$\Lambda$ [$10^6\,$ GeV]}}
\put(258,240){{$d_{Tl}(d_e(y_h))$}}
\end{picture}
 \caption{\footnotesize Constraints on the benchmark scenarios from contributons to $d_{Tl}(d_e)$.}
\label{b3} 
\end{figure}
\noindent
The results are shown in Figs.~(\ref{b1}-\ref{b3}), with the observables normalized to their current experimental bound plotted
against the threshold scale $\La$.  All of the scenarios exhibit broadly similar sensitivity to the degenerate spectrum
utilized previously. However, there is still significant variation in terms of the level of sensitivity exhibited within different benchmark
spectra. 

If we focus first on Fig.~\ref{b1}, the constraints on $C_S$  and $\bar{g}_{\pi NN}$ are most 
stringent within SPS1a, while SPS2 exhibits least sensitivity simply through having a generically heavy SUSY 
spectrum. 

Fig.~\ref{b2} shows the constraints imposed by lepton flavour violating observables, and the quadratic sensitivity to
$\La$ is clearly evident in comparison to the EDM bounds. The strongest constraints again arise within SPS1a and SPS4 due to
having a generically lighter SUSY spectrum. The differences in the constraints from $\mu\rightarrow e\gamma$ are particularly
marked, with SPS2 and SPS8 exhibiting rather minimal sensitivity. This can be understood from the two-loop 
amplitude as due to the relatively small stop mixing in these cases.

Fig.~\ref{b3} exhibits the constraints from the electron EDM. The constraints from the two-loop amplitude on the left
naturally exhibit very similar features to the constraints from $\mu\rightarrow e \gamma$, given the similarity
between the two dipole amplitudes. The constraints on the Higgs operator on the right of Fig.~\ref{b3} are most pronounced
in SPS4 as one would expect due to $\tan\beta$-enhancement. The weak constraint from SPS2 is primarily because of the
large value of $m_{12}^2$ which acts to suppress the effect of the additive complex shift.

\section{Discussion}

So far we have kept our discussion rather general within the context of effective field theory, without 
concerning ourselves  with the details of particular renormalizable UV models.
In the case of the see-saw operator and proton-decay operators, such 
models are well-studied. We would now like to briefly provide an
example of how the effective terms in the superpotential 
studied in this paper can be generated by renormalizable interactions.

We will limit our discussion here to scenarios in which these operators can be generated
by tree level exchange of additional heavy states. As a basic example, consider the MSSM
with an expanded Higgs sector -- an additional heavy singlet $S$ and heavy pair of doublets $H_u'$ and $H_d'$.
This is sufficient to generate all the operators we have considered assuming renormalizable interactions of
the form:
\ba
{\cal W} &=&  \frac{1}{2}MS^2 + \ka S H_u H_d - \mu H_d H_u
- \mu' H_d' H_u'\\
 &&+ U Q( Y_u  H_u + Y_u'  H_u') -  D Q (Y_d H_d + Y_d'H_d') - E L (Y_e H_d+Y_e' H_d'). \nonumber
\label{4higgs}
\ea
Integrating out the singlet $S$ will clearly generate the operator $(H_uH_d)^2$. 
The complex parameters $\mu$ and $\mu'$ are the eigenvalues of a 2$\times$2
complex matrix of $\mu$ parameters that can always be reduced to diagonal form by 
bi-unitary transformations in the $(H_u,H_u')$ and $(H_d,H_d')$ spaces. Assuming the 
hierarchy, $\mu'\gg \mu$, we can integrate out heavy Higgs superfields, 
producing a set of dimension five operators,
\be
{\cal W}^{(5)} = \frac{\ka^2}{M} H_uH_dH_uH_d + \fr{Y_e'Y_u'}{\mu'}ELUQ + \fr{Y_u'Y_d'}{\mu'}(UQ)(DQ).
\label{4higgseff}
\ee
Comparing (\ref{4higgseff}) with (\ref{qule}), we can make the identification $y_h/\La_h = \ka^2/M$, 
$\tilde Y^{qq} = 0$, $Y^{qe}/\Lambda^{qe} = Y_e'Y_u'/\mu'$,
and $Y^{qq}/\Lambda^{qq} = Y_u'Y_d'/\mu'$ and translate 
the sensitivity to the $\Lambda$'s into a sensitivity 
to the extra Higgs fields. Since {\em a priori} there is no correlation 
or dependence between the two sets of Yukawa matrices, 
one can expect novel flavour and $CP$ violating effects 
induced by (\ref{4higgseff}). More specific predictions 
could be made in models that predict or constrain 
the Yukawa couplings $Y$ and $Y'$, due {\em e.g.} to horizontal flavour 
symmetries, Yukawa unification, or discrete symmetries such as parity or $CP$.  

All the  effects which decouple as $1/\Lambda$, when 
put in the language of  the model (\ref{4higgs}), probe the exchange of 
heavy Dirac fermions, namely Higgsino particles composed from 
$\tilde H_u'$ and $\tilde H_d'$. It is then natural to ask 
the question of whether the dimension six
operators induced by the exchange of the heavy scalar Higgses
could provide better  sensitivity to $\mu'$. It is easy to see that in the 
case of arbitrary $Y'_{u,d} \sim O(1)$, the contribution of
dimension six operators to the $K$ meson mass splitting is
\be
\Delta m_K \sim \fr{0.25 {\rm ~ GeV}^3}{\mu'^2} 
\qquad \Longrightarrow \qquad \mu' \ga 8\times 10^6~{\rm GeV},
\label{dim6dk}
\ee
while $\epsilon_K$ is sensitive to scales $\sim 1 \times 10^8$ GeV. 
The reason why this dimension-six contribution dominates so dramatically over
(\ref{dmklimit}) and (\ref{epsk}) is the suppression of the dimension five 
effects by loop and Yukawa factors (\ref{ydsnumber}).

We conclude that $\Delta F=2$ processes mediated by dimension-six operators
in the MSSM extended by an additional pair of Higgses 
 comes very close in sensitivity to the estimates (\ref{edmqq}) and (\ref{cslimit}),
with the latter being somewhat more stringent. This statement does of course depend on the SUSY mass 
spectrum, and having heavier squarks and gluinos would reduce the EDM sensitivity. 
In contrast we should also note that unlike the previous limits 
(\ref{cslimit}) and (\ref{muelimit}), the constraint (\ref{dim6dk}) is 
essentially `static', {\em i.e.} difficult to improve upon, as there is a limited extent to which new physics
contributions to $\Delta m_K$ and $\epsilon_K$ can be isolated from SM uncertainties.

We will end this section with a few additional remarks on issues that we
touched on in this work:

(i) Thus far, we have studied the subset of all possible dimension 
five operators neglecting, for example, $R$-parity violation. 
It is easy to see, however, that no strong constraints on the $R$-parity violating 
terms in (\ref{dim5rpv}) would arise at dimension-five level. 
Indeed, limits on $R$-parity violation usually come from 
SM processes which have to be bilinear in $R$-parity violating 
parameters. Thus, only a combination of two dimension-five terms, 
or  a dimension-five term with a  dimension-four term,
would induce four-fermion operators for example. Since the dimension-four 
terms are tightly constrained (see, {\em e.g.} \cite{sakis}), 
one would not expect the limits on dimension-five operators to be competitive 
with (\ref{cslimit}). 

(ii) A primary goal of any theory of $CP$ violation
is to provide a solution to the strong $CP$ problem. We have shown that the 
effective shift in $\theta$ can be quite significant even if 
higher-dimensional operators are suppressed by $10^{17}$ GeV. 
This has implications for  solutions to the strong $CP$ problem that do not employ 
the dynamical relaxation of $\bar\theta$. For example, if
$\bar \theta =0$ is achieved due to a new global symmetry 
that forces $m_u = 0$ at the dimension-four level but is broken {\em e.g.} by
quantum gravity effects, one could expect the emergence of Planck-scale 
suppressed operators in the superpotential and, remarkably enough, progress in neutron EDM measurements 
by just one or two orders of magnitude would directly probe such a scenario!
Similarly, supersymmetric models that construct a small $\bar\theta$ using 
discrete symmetries can also be affected by these operators,
with possible observable consequences for the neutron EDM. 

(iii) Since the $CP$-odd effective interaction $C_S\bar NN \bar e i \gamma_5 e$ 
provides the leading sensitivity to the energy scale of new physics 
encoded in the semileptonic dimension-five operator in the superpotential, 
it is prudent to recall that  the best constraint on $C_S$ comes from the EDM of the Tl atom, which is 
also used for extracting a constraint on $d_e$. To make both bounds independent 
of the possibility for mutual cancellations, one should use experimental 
information from other atomic EDM measurements. In this respect, the interpretation of 
promising new molecular EDM experiments that aim to improve the sensitivity  to $d_e$ \cite{YbF,PbO}
will require additional theoretical input on the exact dependence on $C_S$. 

(iv) Finally, we would like to emphasize that the main result of this paper, 
namely the sensitivity to the high-energy scale in Eqs.~(\ref{cslimit}) and (\ref{muelimit}),
is quite robust in the sense that it has a mild dependence on the SUSY threshold as exhibited in 
the preceding section.
For example, an increase of the average superpartner mass to 3 TeV would reduce the sensitivity to 
$\Lambda^{qe}$ and $\Lambda^{qq}$ by only a factor of 10, still probing scales of 
a few$\times 10^{7}$ GeV. Contrary to this, the dependence of the electron EDM on the 
Higgs operator is highly  dependent on the details of the SUSY spectrum, as
taking $\tan\beta \sim 5$ and $m_A\sim M_{\rm SUSY} \sim 1 $TeV would reduce the sensitivity to 
$\Lambda_h$ to a few TeV.

\section{Conclusions}

Continuing progress in precision experiments searching for $CP$- 
and flavour-violation provides an increasingly stringent test for 
models of new physics beyond the electroweak scale, and supersymmetric 
theories in particular.  In this paper, we have presented an analysis of flavour and $CP$
violating effects in a two-stage theoretical framework: assuming first that 
the SM becomes supersymmetric at or near  the weak scale, and then that the MSSM 
gives way at some higher scale $\La$ to a theory with additional degrees of freedom.
If nature indeed chooses supersymmetry, both steps can clearly be justified.
The first one follows from the solution to the gauge 
hierarchy problem offered by SUSY and the evidence for the 
second (third, etc.) energy scale comes from rather intrinsic features that are required by, but not contained within, the MSSM: 
mediation of SUSY breaking, neutrino masses, not to mention problems which require other solutions, i.e.
baryogenesis, the strong $CP$ problem, etc.

We have examined  new flavour- and $CP$-violating effects 
mediated by dimension five operators in the superpotential 
to show that sensitivity to these operators extends far beyond the weak scale, 
and indeed probes very high energies. The semi-leptonic operators that 
mediate flavour-violation in the leptonic sector and/or break $CP$ could be 
detectable even if the scale of new physics is as high as $10^{9}$ GeV. 
Since the effects studied here decouple linearly, 
an increase of sensitivity by just two orders of  
magnitude would already start probing the scales that are relevant for 
Majorana neutrino physics. It is also important to note that theoretically, 
should a major breakthrough in the precision of  EDM measurements take place, 
there is ample room for
the EDMs of paramagnetic atoms to probe $CP$-violating operators 
suppressed by the $GUT$ or string scale without facing the SM background from 
the Kobayashi-Maskawa phase, which is known to induce 
tiny EDMs in the lepton sector \cite{myself}. 

The MSSM can contain a variety of  new sources of flavour- and $CP$-violation 
related to the soft-breaking sector. This plethora of sources appears highly excessive 
given the rather minimalist pattern of $CP$ and flavour violation observed experimentally.
A number of model-building scenarios have addressed this issue, often 
successfully, especially if supersymmetry is broken at a relatively
low energy scale. Supposing that the wish of many theorists is granted, 
and a $CP$-symmetric, and flavour-conserving, pattern of soft SUSY breaking is 
achieved in a compelling manner, we may ask the following question: is there any new information 
about such a SUSY theory  that could be provided by the continuation of the low-energy precision
experimental program? This paper provides a clear  affirmative answer to this question.

\subsection*{Acknowledgements}
This work was supported in part by NSERC, Canada. Research at the Perimeter Institute 
is supported in part by the Government
of Canada through NSERC and by the Province of Ontario through MEDT. 

\pagebreak

\section*{Appendix A}

In this appendix, we summarize the 1-loop renormalization group equations used to evolve the dimension five operators
down to the soft threshold. This evolution of course arises purely from the K\"ahler terms.

In general, we have:
\ba
\frac{d}{dt} y_h &=& y_h \left[ \frac{1}{16 \pi^2} \left( \Tr [ \,  
6 \, y_u  y_u^\dag +
6 \, y_d  y_d^\dag + 2 \, y_e  y_e^\dag \, ] - 6 \, g_2^2 
- \frac{6}{5} g_1^2 \right) + \ldots
\right]   \\
\frac{d}{dt} Y^{qe}_{ijkl} &=& \frac{1}{16 \pi^2} \left[ Y^{qe}_{ijkm} (y_e^\dag
y_e)_{ml} + Y^{qe}_{ijml} (2 \, y_e^\dag y_e)_{mk} 
+ Y^{qe}_{imkl} (y_u^\dag y_u + y_d^\dag y_d)_{mj} \right. \nonumber \\
&& \left. + Y^{qe}_{mjkl} (2 \, y_u^\dag y_u)_{mi} - Y^{qe}_{ijkl} \left(
\frac{16}{3} g_3^2 + 3 \, g_2^2 + \frac{31}{15} g_1^2 \right) \right] + \ldots  \\
\frac{d}{dt} Y^{qq}_{ijkl} &=& \frac{1}{16 \pi^2} \left[ Y^{qe}_{ijkm} (y_u^\dag
y_u + y_d^\dag y_d)_{ml} + Y^{qq}_{ijml} (2 \, y_d^\dag y_d)_{mk} 
\right. \nonumber \\ &&  
+ Y^{qq}_{imkl} (y_u^\dag y_u + y_d^\dag y_d)_{mj} 
+ Y^{qq}_{mjkl} (2 \, y_u^\dag y_u)_{mi} \nonumber \\ &&
\left. - Y^{qq}_{ijkl} \left(
\frac{32}{3} g_3^2 + 3 \, g_2^2 + \frac{11}{15} g_1^2 \right) \right] + \ldots  
\ea
and similarly for $\tilde Y^{qq}_{ijkl}$, where $y_u, y_d, y_e$ are $3 \times 3$ Yukawa matrices, and the dots represent
higher order terms.

In practice, since only the third generation Yukawa couplings are significant, we can make use of 
the simplified RGEs, 
\ba
\frac{d}{dt} y_h &\simeq& \frac{y_h}{16 \pi^2} \left[ 6 \, y_t  y_t^\ast +
6 \, y_b  y_b^\ast + 2 \, y_\tau  y_\tau^\ast 
- 6 \, g_2^2 - \frac{6}{5} g_1^2  
\right]  \\
\frac{d}{dt} Y^{qe}_{uuee} &\simeq& \frac{Y^{qe}_{uuee}}{16 \pi^2} \left[  
- \left( \frac{16}{3} g_3^2 + 3\,  g_2^2 + \frac{31}{15} g_1^2 \right) \right] 
\\
\frac{d}{dt} Y^{qe}_{ttee} &\simeq& \frac{Y^{qe}_{ttee}}{16 \pi^2} \left[ 
3 \, y_t^\ast y_t + y_b^\ast y_b -  
\left( \frac{16}{3} g_3^2 + 3\,  g_2^2 + \frac{31}{15} g_1^2 \right) \right]   
\\
\frac{d}{dt} Y^{qe}_{uue\mu} &\simeq& \frac{Y^{qe}_{uue\mu}}{16 \pi^2} \left[  
- \left( \frac{16}{3} g_3^2 + 3\,  g_2^2 + \frac{31}{15} g_1^2 \right) \right] 
\\
\frac{d}{dt} Y^{qe}_{tte\mu} &\simeq& \frac{Y^{qe}_{tte\mu}}{16 \pi^2} \left[ 
3 \, y_t^\ast y_t + y_b^\ast y_b -  
\left( \frac{16}{3} g_3^2 + 3\,  g_2^2 + \frac{31}{15} g_1^2 \right) \right]   
\\
\frac{d}{dt} Y^{qq}_{uudd} &\simeq& \frac{Y^{qq}_{uudd}}{16 \pi^2} \left[ 
- \left(
\frac{32}{3} g_3^2 + 3 g_2^2 + \frac{11}{15} g_1^2 \right) \right]   
\ea

\pagebreak

\end{document}